\documentclass[preprint,12pt]{elsarticle}

\usepackage{amssymb}
\usepackage{lineno}
\usepackage{algorithm} 
\usepackage{algpseudocode}
\usepackage{csquotes}
\usepackage[acronym, automake]{glossaries}
\usepackage{amsmath}
\usepackage{eqnarray}
\usepackage{tabularray}
\usepackage{color}
\usepackage{colortbl}
\usepackage{booktabs}
\usepackage{subcaption}
\usepackage{url}
\usepackage{breakurl}
\usepackage[table,xcdraw]{xcolor}
\usepackage{placeins}
\usepackage{framed}
\usepackage{natbib}

\makeglossaries

\newcommand{\numberoffarms}{73 }

\begin{document}

\begin{frontmatter}



\title{Seamless short- to mid-term probabilistic wind power forecasting}

\author[inst1]{Gabriel Dantas\corref{cor1}}
\author[inst1]{Jethro Browell}

\cortext[cor1]{Corresponding author}

\affiliation[inst1]{
    organization={School of Mathematics and Statistics, University of Glasgow},
    addressline={132 University Pl}, 
    city={Glasgow},
    postcode={G12 8TA}, 
    country={UK}
}

\begin{abstract}

This paper presents a method for probabilistic wind power forecasting that quantifies and integrates uncertainties from weather forecasts and weather-to-power conversion. By addressing both uncertainty sources, the method achieves state-of-the-art results for lead times of 6 to 162 hours, eliminating the need for separate models for short- and mid-term forecasting. It also improves short-term forecasts during high weather uncertainty periods, which methods based on deterministic weather forecasts fail to capture. The study reveals that weather-to-power uncertainty is more significant for short-term forecasts, while weather forecast uncertainty dominates mid-term forecasts, with the transition point varying between wind farms. Offshore farms typically see this shift at shorter lead times than onshore ones. The findings are supported by an extensive, reproducible case study comprising 73 wind farms in Great Britain over five years.

\end{abstract}











\end{frontmatter}

\newacronym{tso}{TSO}{Transmission System Operator}
\newacronym{nwp}{NWP}{Numerical Weather Prediction}
\newacronym{wtg}{WTG}{Wind Turbine Generator}
\newacronym{emos}{EMOS}{Ensemble Model Output Statistics}
\newacronym{bma}{BMA}{Bayesian Model Averaging}
\newacronym{cdf}{CDF}{Cumulative Distribution Function}
\newacronym{pdf}{PDF}{Probability Density Function}
\newacronym{crps}{CRPS}{Continuous Ranked Probability Score}
\newacronym{gbt}{GBT}{Gradient Boosting Trees}
\newacronym{gpd}{GPD}{Generalized Pareto Distribution}
\newacronym{bmra}{BMRA}{Balancing Mechanism Reporting Agent}
\newacronym{bav}{BAV}{Bid Acceptance Volume}
\newacronym{ml}{ML}{Machine Learning}

\pagebreak
\printglossary[type=\acronymtype, style=long,nonumberlist, title=Acronyms]



\FloatBarrier
\section{Introduction}
\label{sec:Introduction}

Wind power is a key technology in low-carbon power systems \citep{ren21-2022,gielen_et_al-2019}; however, high penetration of wind power presents significant operational challenges for electricity markets and networks \citep{heptonstall_and_gross-2021, petinrin_and_shaaban-2016, djorup_et_al-2018,smith_et_al-2010, moreno_et_al-2012}. Electricity markets and networks were designed for dispatchable generation \citep{petinrin_and_shaaban-2016}, but wind power production is stochastic and intermittent with limited predictability. In this context, wind power forecasting from six hours to seven days ahead is essential for the economic and reliable integration of wind energy into existing markets and networks \citep{sweeney_et_al-2019, notton_et_al-2018, book_morales, ahmed_and_khalid-2019, giebel_and_kariniotakis-2017}. Forecast information for horizons up to forty-eight hours ahead directly influences decisions made by Transmission System Operators (\acrshortpl{tso}), such as unit commitment, congestion management, and scheduling storage in addition to decisions related to energy trading in day-ahead and intraday markets \citep{ahmed_and_khalid-2019, sweeney_et_al-2019}. Forecasts from two to seven days ahead influence operational planning and maintenance scheduling \citep{han_et_al-2019}.

Historically, network operation and market participation have been based on single-point forecasts, i.e., based exclusively on the expected value of the energy generation/demand. However, the way forecast information is used is evolving. Probabilistic forecasts are increasingly utilised directly via visualisation as well as in forecast-based decision-support and automation \citep{bessa_et_al-2017, sweeney_et_al-2019}. Probabilistic forecasting is attractive because it provides information about the expected value plus associated forecast uncertainty.

The methods employed in wind power forecasting intrinsically relate to the time scale of interest. Probabilistic wind power forecasting targeting very short-term horizons (up to six hours ahead) is characterised by time series-based methods \citep{tawn_and_browell-2022, giebel_et_al-2011, sweeney_et_al-2020}. However, the performance of time series-based models degrades rapidly for longer horizons. Therefore, for horizons from six hours to seven days ahead, it is necessary to use Numerical Weather Prediction (\acrshort{nwp}) as inputs to statistical and machine learning (ML) models.

\acrshort{nwp} models forecast atmospheric behaviour by numerically solving a system of non-linear differential equations that describe physical processes after first estimating the current state of the atmosphere. The errors associated with \acrshort{nwp} models are caused by inaccuracies related to subgrid dynamical and physical processes, as well as errors associated with initial and boundary conditions \citep{bauer_et_al-2015}. Understanding such errors is fundamental to interpreting \acrshort{nwp}, as the atmosphere is a chaotic system. Thus, ensemble \acrshort{nwp} models, which produce multiple distinct forecast scenarios based on perturbed initial conditions and model physics, have been developed to quantify \acrshort{nwp} uncertainty.

\acrshort{nwp}-based wind power forecasting may be separated into two main tasks. The first involves correcting the \acrshort{nwp} to obtain a more reliable and accurate description of the atmospheric variables at the location of interest. This correction is necessary, as the \acrshort{nwp} model's intrinsic errors usually culminate in biased and underdispersed predictions when compared to observations made at a particular location \citep{Phipps_et_al-2022, vannitsem_et_al-2021}. The second task is associated with modelling the relationship between atmospheric variables predicted by the \acrshort{nwp} model and the wind farm's power output (here, called the weather-to-power relationship\footnote{The terms wind-to-power \citep{casciaro_et_al-2022}, power curve \citep{pinson_and_madsen-2009}, or simply forecasting model \citep{Phipps_et_al-2022} are usually used in the literature related to wind power forecasting. However, this work uses the term weather-to-power to emphasise that atmospheric variables other than wind speed are commonly used as input for this modelling task.}).

NWP-based methods are used to forecast output wind power for a wide range of horizons with strong use-cases for \acrshort{tso}s and energy markets from six-hours to seven-days ahead. Usually, these forecast horizons are subdivided into short- or mid-term, where users reasonably expect different levels of forecast performance. However, the wind power forecasting literature lacks a well-defined threshold between these time scales. Furthermore, while there is an emerging consensus on state-of-the-art methods for short-term forecasting, this is lacking for mid-term. Thus, in many cases, classification developed within the scope of wind speed forecasting is used.

While the classification of NWP-based forecasting methodologies according to short- and mid-term it not well defined, the type of information provided by the \acrshort{nwp} model is a differentiating characteristic related to forecast horizon. NWP-based methodologies can be classified by being based on deterministic or ensemble \acrshort{nwp} models. Deterministic \acrshort{nwp} models provide a single-valued forecast of an atmospheric quantity at a specific location. On the other hand, ensemble \acrshort{nwp} models run many similar but not identical versions of the \acrshort{nwp} model in parallel (e.g., different initial conditions and model parameterisations). Therefore, ensemble \acrshort{nwp} models provide multiple scenarios for the same atmospheric quantity and location, allowing the quantification of \acrshort{nwp} model uncertainty \citep{bauer_et_al-2015}.

The following sub-sections summarise the state-of-the-art NWP-based probabilistic wind power forecasting methodologies powered by deterministic \acrshort{nwp} and ensemble \acrshort{nwp}, respectively. Finally, we present the contributions of this work.




\subsection{Methodologies based on deterministic NWP}
\label{sec:intro_det}

Short-term wind power forecasting methodologies based on deterministic \acrshort{nwp} are not provided with direct quantification of weather forecast uncertainty. Thus, all uncertainty estimation is performed by post-processing. These methods do not differentiate between \acrshort{nwp} bias correction and weather-to-power modelling:  statistical or \acrshort{ml} models directly predict wind farm power output from deterministic \acrshort{nwp} model outputs.  

The statistical and \acrshort{ml} models employed by these methodologies can be divided into models that produce parametric or non-parametric density forecasts. Parametric forecasts assume that the distribution of the predicted wind power at a given time follows a given probability distribution, e.g. Beta, see for example \citet{lange-2005, bludszuweit_et_al-2008, pinson-2012, bremnes-2006, messner_et_al-2014}. In general, parametric methods are less computationally demanding than non-parametric. However, the use of parametric forecasts has limitations. First, the challenge related to choosing a family of appropriate distributions to describe wind power generation. Furthermore, due to the complexity of wind power production, the family of distributions that best describes wind power can change over time and between wind farms. Therefore, parametric forecasts are restrictive and may be too inflexible to produce accurate forecasts in many settings \citep{bessa_et_al-2017}.

As a result, non-parametric wind power forecasting methodologies have been established as best-in-class in recent literature and forecasting competitions for horizons up to two days-ahead \citep{hong_et_al-2016, bessa_et_al-2017, sweeney_et_al-2019}. These include linear quantile regression (e.g., \citet{juban_et_al-2016}); gradient boosting trees (\acrshort{gbt} - e.g., \citet{andrade_and_bessa-2017}, \citet{nagy_et_al-2016}, and \citet{landry_et_al-2016}); kernel density estimation (e.g., \citet{guan_et_al-2020, juban_et_al-2007, zhang_and_wang-2015, dong_et_al-2022, bessa_et_al-2012}), and analogues (e.g., \citet{junk_et_al-2015, alessandrini_et_al-2015, shahriari_et_al-2020, mangalova_and_shesterneva-2016, zhang_and_wang-2015}). In addition, many of the methods proposed in the recent literature use feature engineering to pre-process \acrshort{nwp} model outputs to extract temporal information, spatial information, or both, from multi-dimensional \acrshort{nwp} data (e.g., \citet{andrade_and_bessa-2017, juban_et_al-2016, nagy_et_al-2016}). Feature engineering plays a crucial role in maximising performance, as evidenced in forecast competitions. Results indicate that feature engineering can impact forecast performance as much as the choice of statistical/\acrshort{ml} model \citep{hong_et_al-2016, andrade_and_bessa-2017}.

\subsection{Methodologies based on ensemble NWP}
\label{sec:intro_ens}

Ensemble \acrshort{nwp} models provide multiple scenarios for the same atmospheric quantity in a specific spatial position called ensemble members. Thus, they allow the estimation of weather forecast uncertainty as well as the most likely outcome \citep{bauer_et_al-2015}. The literature on methodologies that leverage ensemble \acrshort{nwp} is much smaller than that of deterministic \acrshort{nwp}. The limited number of related works is perhaps due to the size, complexity and accessibility of ensemble datasets, though these barriers are reducing. Furthermore, most evaluate overall forecast performance in terms of average metrics and overlook the potential of ensemble information in high-impact situations, such as periods of high weather uncertainty, as we explore in Section \ref{sec:results_probw2p_high_unc}.

Forecasting methods based on ensemble \acrshort{nwp} can be classified according to the number of stages used to model the target, i.e., single- and dual-stage methodologies. In dual-stage methodologies, one stage is dedicated to weather-to-power modelling and the other to post-processing the resulting ensemble of power forecasts. Using a dedicated weather-to-power model allows dual-stage methodologies to use established ensemble post-processing methods in the second stage to produce the final wind power forecast. Models originally proposed for atmospheric variables are generally unable to explain the non-linear, bounded, weather-to-power relationship skillfully. Single-stage methodologies use only one stage to forecast the output wind power based on the information in the \acrshort{nwp} ensemble.

Single-stage methodologies usually use more complex \acrshort{ml} and statistical models than dual-stage ones. The models used in single-stage methodologies are often non-linear and provide less interpretability\footnote{Interpretability can be understood in this context as ``the degree to which an observer can understand the cause of a decision'' \citep{miller-2019}}. Only a small number of studies propose single-stage methodologies. Among the main contributions are \citet{wu_et_al-2018}, proposing Lower Upper Bound Estimation with shallow neural networks, \citet{zhang_et_al-2021} proposed a deep learning framework based on a multi-source temporal attention network; and \citet{fujimoto_et_al-2023} proposed using natural gradient boosting for probability density prediction. However, they are all complex methods that demand high computational effort and allow low interpretability. Furthermore, some do not follow best practices for forecast evaluation, restricting their informativeness.

Regarding dual-stage methodologies, the stage related to weather-to-power modelling is usually performed by statistical and \acrshort{ml} models. The models used to describe the weather-to-power relationship can be linear (e.g., polynomial regression \citep{kim_and_hur-2018}, segmented linear regression \citep{pinson_and_madsen-2009, Phipps_et_al-2022}) or non-linear (e.g., logistic regression \citep{nielsen_et_al-2004}, artificial neural networks \citep{junk_et_al-2012, Phipps_et_al-2022}, and random forests \citep{Phipps_et_al-2022}). The stage related to ensemble calibration generally uses statistical models originally proposed to forecast atmospheric variables. 

Statistical models used in this context are typically easy to implement, require low computational effort and are relatively interpretable. The most prominent statistical models used for ensemble calibration are Bayesian Model Averaging (\acrshort{bma}) (e.g., \citep{pinson_and_madsen-2009}) and Ensemble Model Output Statistics (\acrshort{emos})\footnote{\acrshort{emos} is also referred to in the literature as nonhomogeneous regression.} \citep{book_gneiting} (e.g., \citep{Phipps_et_al-2022}). \acrshort{bma} was originally proposed in \citet{raftery_et_al-2005} in order to re-calibrate ensemble forecasts. It can be understood as a mixture of parametric distributions, i.e., kernels. In \acrshort{bma}, each \acrshort{nwp} model ensemble member is dressed with a probability distribution. The \acrshort{bma} coefficients control each kernel's weight and variance and are estimated by maximising some likelihood. Therefore, \acrshort{bma} can be understood as a non-parametric model. On the other hand, \acrshort{emos}, originally proposed in \citet{gneiting_et_al-2005}, is a parametric model widely used in the post-processing of \acrshort{nwp} ensembles. \acrshort{emos} aims to describe the posterior distribution parameters as functions of ensemble statistics (e.g., the average and standard deviation of ensemble members). Due to its computational simplicity and satisfactory results, \acrshort{emos} is widely used in operational post-processing at weather services \citep{Phipps_et_al-2022}. The preference for using \acrshort{emos} also extends to the context of probabilistic wind power forecasting powered by ensemble \acrshort{nwp}. The most commonly used distribution families in \acrshort{emos} applications for wind power are Truncated Normal \citep{Phipps_et_al-2022, gneiting_et_al-2014} and Gamma \citep{Phipps_et_al-2022}. Other models have been developed to calibrate ensembles to forecast atmospheric variables, e.g., member-by-member post-processing \citep{van_schaeybroeck_and_vannitsem-2015} and isotonic distributional regression \citep{henzi_et_al-2021}. However, they have yet to be widely explored in the context of wind power forecasting. Despite the growing literature associated with probabilistic weather-to-power modelling \citep{bilendo_et_al-2023}, dual-stage forecasting methodologies only adopted single-point models, to the best of the authors knowledge. Thus, methodologies powered by ensemble\acrshort{nwp} proposed to date ignore a key source of uncertainty and naively attempt to account for it in ensemble post-processing.

\subsection{Contributions}
\label{sec:intro_contribution}

Methodologies based on deterministic \acrshort{nwp} perform all uncertainty estimation through post-processing. Existing methodologies based on ensemble \acrshort{nwp} either combine ensemble calibration with a deterministic weather-to-power model or forecast power production directly based on features derived from ensemble \acrshort{nwp}. Existing methodologies do not distinguish between sources of uncertainty and implicitly attempt to correct this via post-processing. To the best of the authors' knowledge, this work is the first to propose a method that explicitly models and combines \acrshort{nwp} forecast uncertainty and uncertainty arising from weather-to-power conversion. 

This work's first contribution is a novel method for seamless short- to mid-term wind power forecasting based on probabilistic weather-to-power modelling ensemble post-processing. This method explicitly models and combines the uncertainties from \acrshort{nwp} and the weather-to-power relationship. Through an extensive and fully reproducible case study, the proposed method is shown to have significant performance benefits.

The second contribution is a study of the relative performance of state-of-the-art methods based on deterministic and ensemble \acrshort{nwp}, which is absent from the current literature.

Finally, the lead-time at which ensemble \acrshort{nwp} adds benefit for wind power forecasting is investiagted and shown to vary between wind farms; however, there is a clear difference between onshore and offshore wind farms in Great Britain.

The remainder of this work is organised as follows. Section \ref{sec:methodology} describes the proposed method and the state-of-the-art reference methodologies adopted to compare with the proposed method. Section \ref{sec:studycase} describes the \acrshort{nwp} models and observational data sets. This work used five years of observational data from \numberoffarms wind farms in Great Britain and deterministic and ensemble \acrshort{nwp} models provided by ECMWF. Section \ref{sec:results} presents and discusses the results. Section \ref{sec:conclusion} addresses the conclusions and perspective regarding future works.

\section{Methodology}
\label{sec:methodology}

The method proposed in this work is based on two stages. A probabilistic description of the weather-to-power relationship and the combination of the probabilistic weather-to-power model with the ensemble \acrshort{nwp}. In the following text, $\hat{x}_{t+k|t}$ is the forecast of $x_{t+k}$ produced using the \acrshort{nwp} with base time $t$ (i.e., horizon $k$ steps ahead of the base time).

\subsection{Probabilistic weather-to-power modelling}
\label{sec:methodology_a2p}

Weather-to-power relationship modelling aims to describe the wind farm's output power given actual weather conditions as described by the \acrshort{nwp} model to be used in forecasting. The weather-to-power relationship is uncertain, hence the probability distribution of power production is predicted given certain weather conditions.

Quantile regression with Gradient-Boosted Trees (GBT) is used to model the weather-to-power relationship, which is established as best-in-class for similar tasks \citep{andrade_and_bessa-2017, landry_et_al-2016, gilbert_et_al-2018}. \acrshort{gbt} are based on a non-parametric and non-linear ensemble learning algorithm. They can be understood as ``an ensemble of weakly predictive regression trees, combined to generate a powerfully predictive collective'' \citep{browell_et_al-2020}. However, unlike models, such as random forests, which combine models in the ensemble via averaging, \acrshort{gbt} is based on sequentially adding weak learners (i.e., regression trees) to the ensemble. A new base-learn is trained at each iteration, considering the error of the ensemble created so far. Furthermore, it is very flexible, accommodating a wide range of loss functions (though some implementations require a twice-differentiable loss, which excludes the standard quantile loss) and combinations of continuous and discrete features. See \citet{friedman-2001}, \citet{ke_et_al-2017} and \citet{chen_and_guestrin-2016} for more details.

A separate \acrshort{gbt} is estimated for each of the quantiles of interest. The \acrshort{gbt} hyperparameters are tuned for each quantile using the Bayesian Optimization algorithm with random permutation cross-validation \citep{snoek_et_al-2012}. After choosing a set of hyperparameters, the \acrshort{gbt} is estimated using all the data samples available for weather-to-power model estimation. As different \acrshort{gbt}s are estimated for each quantile, it is not guaranteed that the predicted quantiles will be monotonically increasing (i.e., quantile crossing). Several strategies have been proposed to solve the quantile crossing problem \citep{chernozhukov_et_al-2010, zhou_et_al-2020}. Here, quantiles are sorted to be monotonically increasing  \citep{juban_et_al-2016}. This strategy is adopted due to its simplicity and the satisfactory results achieved.

Weather-to-power modelling aims to predict the wind power given actual atmospheric conditions, as described by the \acrshort{nwp} model. Therefore, the impact of \acrshort{nwp} forecast uncertainty on the weather-to-power model estimate should be minimised. Therefore, only \acrshort{nwp} operational analysis is used to fit \acrshort{gbt}s. The ensemble \acrshort{nwp} in this study provides equally likely future scenarios with no distinguishable features or ordering, i.e., exchangeable ensemble members. Exchangeability implies independence between forecasts performed at different base times. Therefore, estimating distinct predictive models for each ensemble member is not meaningful. Thus, the weather-to-power model is estimated using the ensemble median for each atmospheric variable and grid point. Note that the proposed method was developed for exchangeable ensemble members. However, if non-exchangeable ensembles are available, alternative treatments could be considered. In the following text, $\hat{q}^{(\tau)}_{j, t+k|t}$ is the quantile $\tau$ predicted by ensemble member $j$ from the weather-to-power model. 

\subsection{Ensemble combination}
\label{sec:methodology_combination}

The second stage of the proposed method incorporates weather forecast uncertainty using ensemble \acrshort{nwp}.
The ensemble captures weather forecast uncertainty, and each ensemble member is dressed with a kernel quantifying additional weather-to-power uncertainty for that ensemble member. The parameters of the kernels are derived from the probabilistic weather-to-power predictions described above.
%
%
The mean of the $j$th kernel is given by the median of the weather-to-power predicted for ensemble member $j$ at the corresponding time $t+k|t$
\\
\begin{equation}
    \label{eq:kernel_mean}
    \hat{\mu}_{j, t+k|t} = \hat{q}^{(50)}_{j,t+k|t} \text{ for } j=1, ..., m
\end{equation}
\\
and the standard deviation is a linear function of the interquartile range (IQR)

\begin{equation}
    \label{eq:kernel_std}
    \hat{\sigma}_{j, t+k|t} = \lambda_{0, k} + \lambda_{1, k} \left( \hat{q}^{(75)}_{j, t+k|t} - \hat{q}^{(25)}_{j, t+k|t} \right) \text{ for } j=1, \dots, m
\end{equation}
\\
where $\lambda_{0,k}>0$ and $\lambda_{1,k} \ge 0$ are the spread coefficients related to horizon $k$. The spread coefficients allow variation in the kernel dispersion over the forecast horizons. These variations are expected as the \acrshort{nwp} accuracy tends to degrade as the forecast horizon increases, though this does not adjust the dispersion of the ensemble itself. 

Using parametric kernels allows the Cumulative Distribution Function (CDF) associated with each ensemble member to be described by just three quantiles, significantly reducing computational effort compared to multiple quantile regression. In this work, the kernels follow a Normal distribution with \acrshort{cdf} given by
\\
\begin{equation}
    \label{eq:kernel_normal}
    \hat{F}_{j, t+k|t}(x) = \Phi\left(x; \hat{\mu}_{j, t+k|t}, \hat{\sigma}_{j, t+k|t} \right) \text{ for } j=1, ..., m
\end{equation}
\\
where $\hat{F}_{j, t+k|t}$ denotes the estimated \acrshort{cdf}/kernel related to ensemble member $j$ and $\Phi$ is the Normal \acrshort{cdf}. The power data are non-negative, favouring the use of the Gamma distribution. However, both Gamma and Normal kernels produced almost identical results. Therefore, the Normal distribution was chosen for simplicity.



%

A beta-transformed linear opinion pool performs kernel combination to produce the final predictive density. \citet{gneiting_and_ranjan-2013} proposed the beta-transformed linear opinion pool to correct the miscalibration that typically results from simple linear combination due to dependency between kernels that, in many cases, is difficult to estimate. The \acrshort{cdf} resulting from the kernel combination $\hat{G}_{j, t+k|t}$ at $x \in \mathbb{R}$ is given by
\\
\begin{equation}
    \label{eq:blp}
    \hat{G}_{t+k|t}(x) = I_{a_{k},b_{k}}  \left( \frac{1}{m} \sum_{j=1}^{m}  \hat{F}_{j, t+k|t}(x) \right)
\end{equation}

\begin{equation*}
    \label{eq:rbf}
    I_{a_{k},b_{k}}(z) = \frac{B_{a_{k},b_{k}}(z)}{B_{a_{k},b_{k}}} 
\end{equation*}
\\
where $I_{a_{k},b_{k}}$ denotes the \acrshort{cdf} of the Beta density\footnote{The cumulative distribution function of the Beta density is also known as regularised incomplete beta function.} with parameters $a_{k}>0$ and $b_{k}>0$, and $B_{a_{k},b_{k}}(z)$ and $B_{a_{k},b_{k}}$ are incomplete and complete beta functions, respectively.

\begin{figure}
\centering
\includegraphics[scale=0.52]{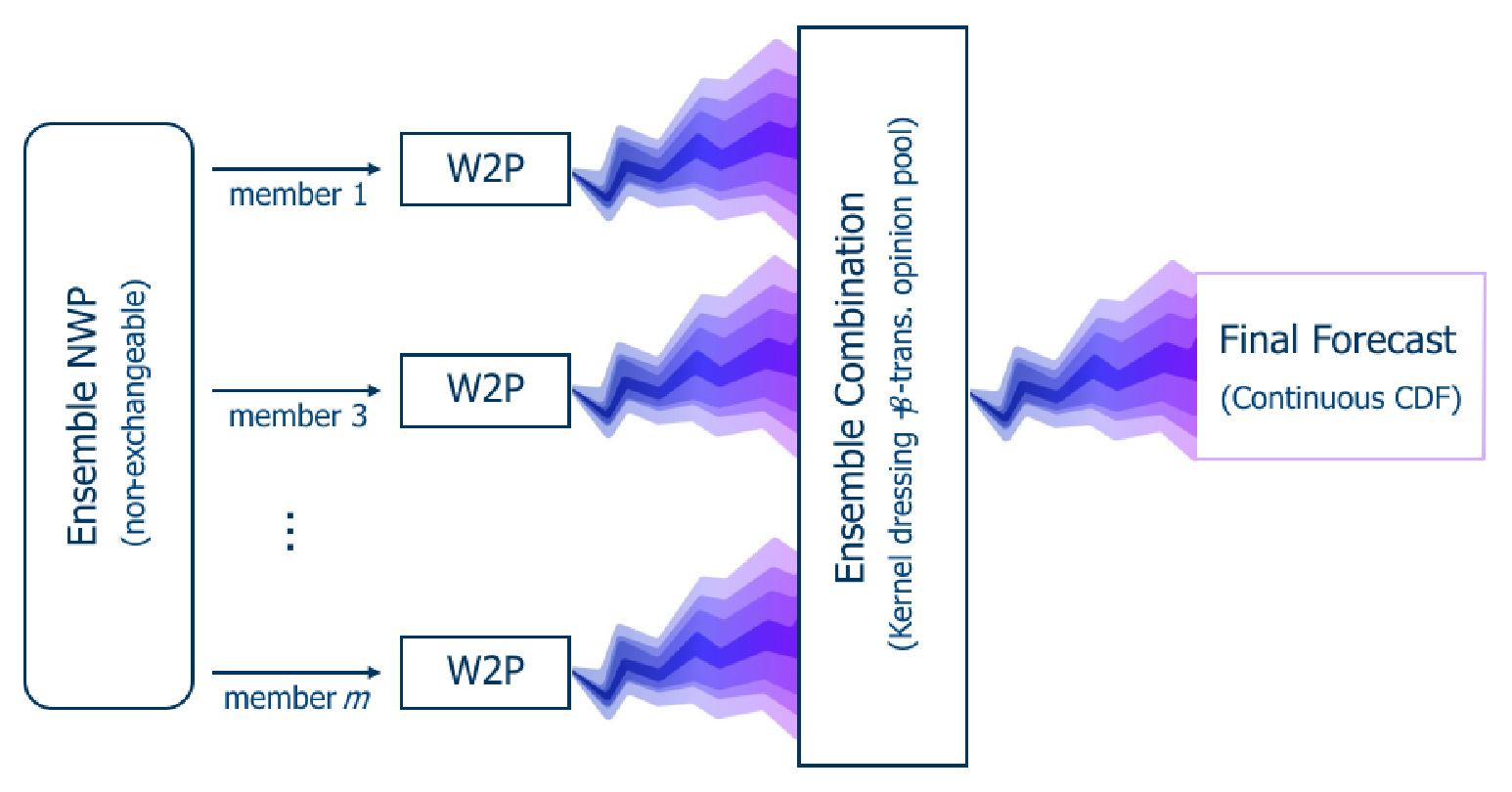}
\caption{Diagram illustrating the proposed probabilistic wind power forecasting. Ensemble of numerical weather predictions are the input. Each ensemble member is converted into a probabilistic wind power prediction using a weather-to-power model. The resulting ensemble of power predictions are combined to produce the final probabilistic wind power forecast.}
\label{Fig:methodology_diagram}
\end{figure}

Four coefficients describe the ensemble combination: two related to the ensemble spread correction, $\lambda_{0,k}$ and $\lambda_{1,k}$, and two associated with the beta-transformed linear opinion pool, $a_k$ and $b_k$. Estimation is performed for each forecast horizon $k$ by minimising a loss function $\ell$ that reflects forecast performance
\\
\begin{equation}
    \label{eq:coef_estimation}
    \left[ \hat{\lambda}_{0,k},\hat{\lambda}_{1,k}, \hat{a}_k, \hat{b}_k \right] = \underset{\lambda_{0,k}, \lambda_{1,k}, a_k ,b_k}{\arg\min} \, \ell\left( \lambda_{0,k}, \lambda_{1,k}, a_k ,b_k \right) \quad .
\end{equation}
\\
The loss function used in this work is the mean Continuous Ranked Probability Score (\acrshort{crps}). The \acrshort{crps} is a strictly proper scoring rule widely used for evaluating probabilistic predictions \citep{gneiting_and_raftery-2007}. It is recommended for probabilistic forecasting assessment because it reflects the paradigm of maximising the sharpness of the predictive distributions subject to calibration, though calibration should still be verified separately \citep{gneiting_et_al-2007}. \acrshort{crps} compares the predicted \acrshort{cdf} $\hat{F}_t$ on $\mathbb{R}$ and the observation $y_t \in \mathbb{R}$. The mean \acrshort{crps} is defined as
\\
\begin{equation}
    \label{eq:crps}
    \ell(\hat{F},y) = \frac{1}{n} \sum_{t=1}^{n} \int_{\mathbb{R}} \left( \hat{F}_t(x) - H(x-y_t) \right)^2 \,dx
\end{equation}
\\
where $y_t$ is the observed value in $t$, $\hat{F}_t$ is the predicted \acrshort{cdf} that depends on $\hat{\lambda}_0$, $\hat{\lambda}_1$, $\hat{a}_k$ and $\hat{b}_k$, and $H(\cdot)$ is the Heaviside step function
\\
\begin{equation*}
    H(z) = 
    \begin{cases}
      0 & \text{$if \;\; z < 0$}\\
      1 & \text{$if \;\; z \geq 0$}\\
    \end{cases}  \quad .
\end{equation*}
\\
Figure \ref{Fig:methodology_diagram} summarises the proposed method.

\subsection{Reference methods}
\label{sec:methodology_refmodels}

This work adopts four reference methods for two main reasons. First, to compare the proposed method with the established state-of-the-art. Second, to investigate the horizon where switching from deterministic to ensemble \acrshort{nwp} adds values to the forecast, i.e., the threshold between short- and mid-term temporal scales.The following subsections present the reference methods, and Table \ref{tab:ref_methods} summarises their main characteristics. The reference methods are named according to the \acrshort{nwp} model used as input, the weather-to-power model, and the ensemble combination/calibration model.

\definecolor{Alto}{rgb}{0.839,0.839,0.839}
\begin{table}
\centering
\caption{Main characteristics of the reference methods and proposed method. Note that the method ENS-QGBT-$\beta$MM refers to the proposed method.}
\label{tab:ref_methods}
\scalebox{0.66}{\begin{tblr}{
  cells = {c},
  row{3} = {Alto},
  row{5} = {Alto},
  vline{2-5} = {-}{},
  hline{1,7} = {-}{0.08em},
  hline{2} = {1}{l},
  hline{2} = {2-4}{},
  hline{2} = {5}{r},
}
\textbf{Method} & \textbf{NWP model} & \textbf{Weather-to-power model} & {\textbf{Calibration/}\\\textbf{ combination model}} & \textbf{Tail model}\\
ENS-GBT-None & {Ensemble \acrshort{nwp}\\ (ECMWF-ENS)} & {Single-value GBT\\(i.e., $\hat{q}_{50}$)} & None & {Peaks Over\\ Thresholds}\\
ENS-QGBT-None & {Ensemble \acrshort{nwp}\\ (ECMWF-ENS)} & {Quantile GBT \\ (see subsection \ref{sec:methodology_a2p})} & None & {Peaks Over\\ Thresholds}\\
HRES-QGBT-None & {Deterministic \acrshort{nwp}\\ (ECMWF-HRES)} & {Quantile GBT\\ originally proposed\\ by \citet{andrade_and_bessa-2017}} & None & {Peaks Over\\ Thresholds}\\
ENS-GBT-EMOS & {Ensemble \acrshort{nwp}\\ (ECMWF-ENS)} & {Single-value GBT\\(i.e., $\hat{q}_{50}$)} & \acrshort{emos} & None\\
ENS-QGBT-$\beta$MM & {Ensemble \acrshort{nwp}\\ (ECMWF-ENS)} & {Quantile GBT\\ (see subsection \ref{sec:methodology_a2p})} & {Mixed model based\\ on kernel dressing and\\ beta-transformed\\ linear opinion pool\\ (see subsection \ref{sec:methodology_combination})} & None
\end{tblr}}
\end{table}

\subsubsection{ENS-GBT-None}
\label{sec:methodology_refmodels_ENS-GBT-None}

The ENS-GBT-None is included as a benchmark. Each ensemble member is converted to power using the median provided by the weather-to-power model (i.e., $q^{(50)}$), no ensemble post-processing is performed before or after weather-to-power conversion. Density forecasts are constructed considering the ensemble members to be quantiles that are interpolated, and with a parametric tail distribution to extrapolate beyond the upper and lowermost quantiles. The parametric distribution used to model the tails is described in subsection \ref{sec:methodology_refmodels_tailsQT}.


\subsubsection{ENS-QGBT-None}
\label{sec:methodology_refmodels_ENS-QGBT-None}

The ENS-QGBT-None is proposed here as the second benchmark. It aims to represent the accuracy obtained using a probabilistic weather-to-power model without performing ensemble correction or calibration. This method is fed by ensemble \acrshort{nwp}. The weather-to-power model used by ENS-QGBT-None is the same as the method proposed in this work; however, no ensemble correction or calibration is performed. Thus, the final forecast is given by
\\
\begin{equation}
    \label{eq:ENS-QGBT-None}
    \tilde{q}^{(\tau)} = \frac{1}{m} \sum_{j=1}^{m} \hat{q}^{(\tau)}_j
\end{equation}
\\
where $\tilde{q}^{(\tau)}$ is the final forecast for the quantile $\tau$, and $\hat{q}^{(\tau)}_j$ is the quantile $\tau$ estimated by the weather-to-power model using as input the ensemble member $j$.

\subsubsection{HRES-QGBT-None}
\label{sec:methodology_refmodels_HRES-QGBT-None}

This reference method was originally proposed by \citet{andrade_and_bessa-2017} and represents the state-of-the-art for short-term wind power forecasting. It was developed by improving the models that obtained first and second place in the Global Energy Forecasting Competition 2014 (GEFCom2014) \citep{hong_et_al-2016}, and Hybrid Energy Forecasting and Trading Competition 2024 \citep{browell_et_al-2023}.

This method is fed by a deterministic \acrshort{nwp} model (in this case, ECMWF HRES) and produces non-parametric density forecasts. Feature engineering is used to extract spatial and temporal features from the gridded \acrshort{nwp} data. For this method only, separate \acrshort{gbt}s are estimated for each forecast horizon and quantile of interest. The hyperparameter tuning was also done separately for each forecast horizon and quantile using the Bayesian Optimization Algorithm \citep{snoek_et_al-2012}.

\subsubsection{ENS-GBT-EMOS}
\label{sec:methodology_refmodels_ENS-GBT-EMOS}

This reference method is based on the post-processing strategy that achieved the lowest \acrshort{crps} in \citet{Phipps_et_al-2022}. This method represents the state-of-the-art probabilistic wind power forecasting based on ensemble \acrshort{nwp}. It is a dual-stage method, where each \acrshort{nwp} ensemble member is converted into wind power using a single-value weather-to-power model. The wind power ensemble is post-processed using \acrshort{emos} to provide the final probabilistic wind power forecasting.

The weather-to-power model used by ENS-GBT-EMOS is the same as the proposed method. Therefore, it uses the same \acrshort{nwp} input variables and horizontal domain. However, here, the weather-to-power model provides only single-value predictions.

EMOS performs ensemble combination on the ensemble of power forecasts. As discussed in subsection \ref{sec:intro_ens}, \acrshort{emos} is a parametric model widely used in post-processing \acrshort{nwp} ensembles. \acrshort{emos} estimates the parameters of the output density forecast as a function of the median and variance of the ensemble members. Gamma and Truncated Normal distribution families were initially tested, with Gamma achieving lower \acrshort{crps}. Therefore, this reference method adopts the Gamma distribution. Thus, the parameters of the output density forecast are given by
\\
\begin{equation}
    \label{eq:emos_gamma-shape}
    \hat{\alpha}_{t+k|t} = \frac{\hat{\mu}_{t+k|t}^2}{\hat{\sigma}_{t+k|t}^2}
\end{equation}

\begin{equation}
    \label{eq:emos_gamma-scale}
    \hat{\beta}_{t+k|t} = \frac{\hat{\mu}_{t+k|t}}{\hat{\sigma}_{t+k|t}^2}
\end{equation}

\begin{equation}
    \hat{\mu}_{t+k|t} = c_{0,k} + c_{1,k} \overline{\hat{q}^{(50)}}_{t+k|t}
\end{equation}

\begin{equation}
    \hat{\sigma}_{t+k|t} = c_{2,k} + c_{3,k} \sigma_{\hat{q}^{(50)}_{t+k|t}}
\end{equation}
\\
where $\hat{\alpha}_{t+k|t}$ and $\hat{\beta}_{t+k|t}$ are the estimated shape and rate parameter of the Gamma distribution, respectively; $\hat{\mu}_{t+k|t}$ and $\hat{\sigma}_{t+k|t}$ are the estimated mean and standard deviation of the Gamma distribution; $c_{0,k}$, $c_{1,k}$, $c_{2,k}$ and $c_{3,k}$ are the \acrshort{emos} coefficients; and $\overline{\hat{q}^{(50)}}_{t+k|t}$ and $\sigma_{\hat{q}^{(50)}_{t+k|t}}$ are the mean and standard deviation of the output power provided by the ensemble members, respectively. The \acrshort{emos} coefficients are estimated by minimising the \acrshort{crps}. For the \acrshort{emos} coefficients estimation, this work utilises the \acrshort{crps} closed form for Gamma distribution \citep{scheuerer_and_moller-2015}, given by
\\
\begin{eqnarray}
    \label{eq:crps_gamma}
    \text{CRPS}(F_{\mathcal{G}(\alpha,\beta)},x) &= 
        \begin{aligned}[t]
            &x\left( 2 F_{\mathcal{G}(\alpha,\beta)}(x) -1 \right) \\
            &- \frac{\alpha}{\beta} \left(2 F_{\mathcal{G}(\alpha+1,\beta)}(x) -1  \right)\\
            &- \frac{\alpha}{\beta\pi} B \left( \alpha+\frac{1}{2}, \frac{1}{2} \right)
        \end{aligned}
\end{eqnarray}
\\
where $F_{\mathcal{G}(\alpha,\beta)}$ is the \acrshort{cdf} of the Gamma distribution with shape parameter $\alpha$ and rate parameter $\beta$, and $B$ is the complete beta function. Note that the \acrshort{emos} coefficients are estimated separately for each horizon $k$.

\subsubsection{Distribution tails and quantiles of interest}
\label{sec:methodology_refmodels_tailsQT}

The reference methods ENS-GBT-None, HRES-QGBT-None, and HRES-QGBT-None are non-parametric and provide quantiles as forecast output. The quantiles provided by these models range from 5\% to 95\% with 5\% resolution (i.e., 19 quantiles). Methods based on non-parametric models usually present difficulties in describing the tail of the predictive distribution. This difficulty stems from the need for a large volume of data to estimate sporadic events. Thus, this work adopts a hybrid approach using a parametric model to describe the distribution tails similarly to previous works \citep{matos_et_al-2016, pinson_et_al-2009, matos_and_bessa-2011, goncalves_et_al-2021, beirlant_et_al-2017}. The tails are modelled here using the Peak Over Threshold method \citep{mcneil_and_saladin-1997}, where the threshold exceedances follow the Generalized Pareto Distribution for sufficiently extreme thresholds. The Generalized Pareto Distribution (\acrshort{gpd}) \acrshort{cdf} is given by
\\
\begin{align}
    \label{eq:gdp}
    \hat{D}_{t+k|t}(z_{t+k|t};\hat{\eta}_k,\hat{\psi}_k,\hat{\xi}_k) &=
        \begin{cases}
            1 - \left( 1 + \frac{\hat{\xi}_k(z_{t+k|t} - \hat{\eta}_k)}{\hat{\psi}_k} \right)^{-1/\hat{\xi}_k} &\text{ for } \hat{\xi}_k \neq 0 \\
            1 - \exp{\left( \frac{\hat{\eta}_k - z_{t+k|t}}{\hat{\psi}_k} \right)} &\text{ for } \hat{\xi}_k = 0
        \end{cases} \\
        \notag
        z_{t+k|t} &= \left| x_{t+k} - \hat{e}_{t+k|t} \right|
\end{align}
\\
in the above equations, $\hat{D}$ is the estimated \acrshort{gpd}'s \acrshort{cdf}, $z_{t+k|t}$ is the exceedance over the threshold $\hat{e}_{t+k|t}$, $x_{t+k}$ is the output power at ${t+k}$; $\hat{\eta}_k, \hat{\psi}_k>0$ and $\hat{\xi}_k$ are the location, scale and shape parameters estimated for the horizon $k$, respectively. Upper and lower tails can have different behaviours, so the \acrshort{gpd} parameters are independently estimated by maximum likelihood. Here, the threshold for the lower tail is given by the predicted quantile 5\% and the threshold for the upper tail is given by the predicted quantile 95\%. The \acrshort{gpd} is used in this work due to its flexibility, which allows the modelling of extreme events with light bounded tails when $\xi_k<0$, heavy tails when $\xi_k>0$, and exponential distribution when $\xi_k=0$.

The forecast assessment performed in this work demands that forecast outputs be delivered as \acrshort{cdf}s. Using \acrshort{cdf}s in forecast assessment allows a consistent comparison between the different studied methods. Thus, the quantiles provided by the non-parametric reference methods are interpolated to obtain the predicted \acrshort{cdf}. The interpolation is performed using the Piecewise Cubic Hermite Interpolating Polynomial algorithm \citep{fritsch_and_butland-1984} to preserve the monotonicity of the \acrshort{cdf}.

The target variable studied in this work is the wind farm's metered energy output. The data were normalised by each wind farm's available capacity for ease of comparison between wind farms. Thus, the time series used in the context of this work are contained in a range [0,1]. However, the probabilistic forecasts performed by the reference methods are not necessarily constrained to the unit interval. Thus, this work applied the solution proposed by \citet{pinson-2012}. This solution adds support (0,1) and probability masses on the boundaries 0 and 1. Thus, the predicted distributions become contained in the unit interval, maintaining the integral of the predicted \acrshort{pdf} for $x \in [0,1]$ equal to 1. The  Probability Density Function (PDF) with probability masses on the boundaries is defined as
\\
\begin{equation}
    \label{eq:limcorr}
    \hat{f}_{t+k|t}^*(x) = \omega_{0, \,t+k|t}\delta_0 + \hat{f}_{j, t+k|t}(x) + \omega_{1, \,t+k|t}\delta_1 \quad , \quad x \in [0,1]
\end{equation}

\begin{equation*}
    \omega_{0, \,t+k|t} = \hat{F}_{t+k|t}(x=0)
\end{equation*}

\begin{equation*}
    \omega_{1, \,t+k|t} = 1 - \hat{F}_{t+k|t}(x=1)
\end{equation*}
\\
where $\hat{f}_{t+k|t}^*$ is the predicted \acrshort{pdf} with probability masses on the boundaries; $\omega_0$ and $\omega_1$ are the weights related to the probability mass at 0 and 1, respectively; and $\delta_0$ and $\delta_1$ are Dirac delta functions at 0 and 1.

\FloatBarrier
\subsubsection{Assessment framework}
\label{sec:assessment_framework}

This work uses \acrshort{crps}, given by \eqref{eq:crps}, to assess the performance of probabilistic forecasting methods. \acrshort{crps} is a strictly proper scoring rule widely applied to evaluate probabilistic forecasting of continuous variables\citep{messner_et_al-2020}.

CRPS provides a general understanding of the predictive model's accuracy. Therefore, the attributes resulting from the \acrshort{crps} decomposition --- reliability and resolution --- are used to assess specific aspects of the prediction \citep{arnold_et_al-2023, bentzien_and_friederichs-2014}. Reliability concerns the calibration of the forecasting model, i.e., whether the predicted probability agrees with the observed relative frequency. Reliability can be assessed through the value obtained from the \acrshort{crps} decomposition as the resolution or visually through, for example, the reliability diagram \citep{pinson_et_al-2010}. Resolution concerns the ability of the forecasting model to discriminate between different outcomes of an observation.

This work also uses skill scores in order to compare the accuracy of distinct forecasting methods and the significance of any observed differences. The skill score $S$ of the method $s$ with respect to the method $r$ for a performance metric $A$ is given by
\\
\begin{equation}
    \label{eq:skill_score}
    S_{r, s} = \frac{A_r - A_s}{A_r - A_{perf}}
\end{equation}
\\
where $A_r$ is the metric value for the reference method, $A_s$ is the metric value for the method of interest, and $A_{perf}$ is the metric value for the `perfect' prediction. $A_{perf}=0$ for \acrshort{crps}. Additionally, a bootstrap re-sampling of skill scores is used to determine if apparent differences in forecast accuracy (i.e., positive or negative skill score) differ signiﬁcantly from the null hypothesis that skill is zero at the 0.05 level.

The predictive methods studied in this work vary significantly in handling the uncertainty associated with \acrshort{nwp} and weather-to-power. Therefore, analysing the uncertainty associated with each process is necessary to understand predictive methods' capabilities. The \acrshort{nwp} uncertainty assesses the spread of the \acrshort{nwp} ensemble. While the weather-to-power uncertainty is the average spread of the probabilistic power prediction across all ensemble members. Therefore the \acrshort{nwp} uncertainty $U^{(\text{NWP})}$ and the weather-to-power uncertainty $U^{(\text{W2P})}$ are defined as
\\
\begin{equation}
    \label{eq:unc_nwp}
    U_{t+k|t}^{(\text{NWP})} = F^{-1}_{75}\left( \hat{q}^{(50)}_{j,\,t+k|t}  \right) - F^{-1}_{25}\left( \hat{q}^{(50)}_{j,\,t+k|t}  \right) \text{ for } j=1,...,m
\end{equation}

\begin{equation}
    \label{eq:unc_w2p}
    U_{t+k|t}^{(\text{W2P})} = \frac{1}{m} \sum_{j=1}^{m} \left( \hat{q}^{(75)}_{j,\,t+k|t} - \hat{q}^{(25)}_{j,\,t+k|t} \right)
\end{equation}
\\
where $F^{-1}_{\tau}$ is the quantile function for probability $\tau$, and $\hat{q}^{(25)}_{j}$, $\hat{q}^{(50)}_{j}$ and $\hat{q}^{(75)}_{j}$ are the quantiles 25\%, 50\% and 75\% of the estimated output power related to the ensemble member $j$, respectively. Both uncertainty metrics consider the \acrshort{nwp} converted into output power to provide a common framework for uncertainty assessment, facilitating comparison between them.

\section{Case study}
\label{sec:studycase}

This section details the wind power and \acrshort{nwp} datasets, model configurations, and cast study set-up used to evaluate and compare the abovementioned forecasting methods. All data, models and forecasts are shared along with this paper, as described in the Data Availability section.

\subsection{Wind power data}
\label{sec:studycase_obs}

In Great Britain (GB), the Balancing Mechanism Reporting Agent (\acrshort{bmra}) \citep{site_BMRA} is responsible for the settlement of the electricity market. Metered energy output and Bid Acceptance Volume (\acrshort{bav}) time series at 30-minute intervals are used here. The \acrshort{bmra} database is subject to high standards imposed by the energy market \citep{grid_code}.

Wind farms in GB are subject to curtailment. Curtailment can substantially modify the relationship between the weather patterns and the wind farm's power output, significantly affecting forecasting systems \citep{yan_et_al-2022, yan_et_al-2019}. Estimated curtailment volumes are contained in \acrshort{bav} data. Thus, periods with non-zero \acrshort{bav}s were excluded from modelling and forecast evaluation.

The wind power time series were normalised by each wind farm's available (or nominal) capacity to simplify the comparison of performance between wind farms of different capacities. Thus, the normalised time series have range [0,1]. However, the available capacity of wind farms can vary over time. The main reasons for this variation are the addition of new \acrshort{wtg}s to the plant (i.e., during the construction phase) and the unavailability of \acrshort{wtg}s or other electrical equipment due to outages and maintenance. The dataset provided by \acrshort{bmra} does not include information on the farms' available capacity over time. Hence, the method described in the supplementary material was applied to estimate the available capacity at each time stamp (i.e., a time series of available capacity). Thus, this study assumes that the wind farm's capacity is known at the time the forecast is produced.

\begin{figure}[h]
    \centering
    \includegraphics[scale=0.87]{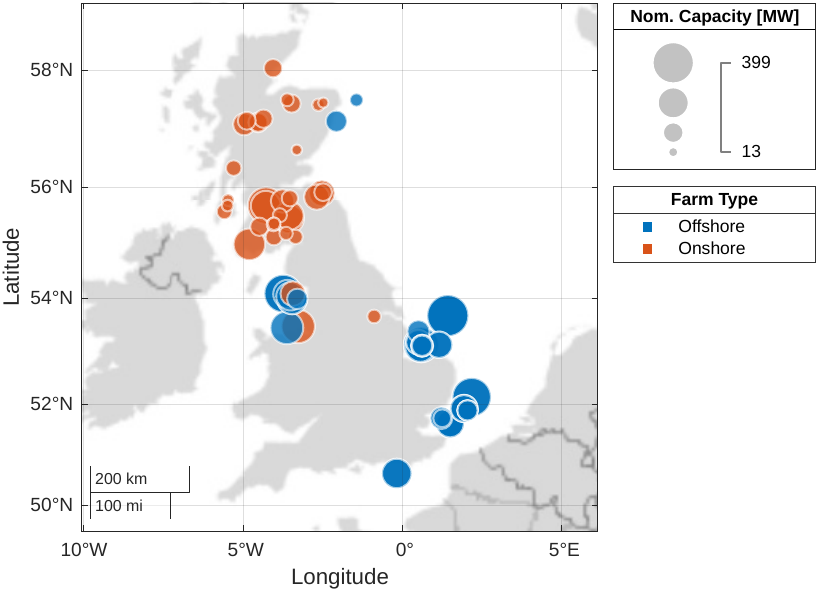}
    \caption{Map showing the wind farms included in the case study. Each wind farm is represented by a circle on the map. The circle's diameter represents the wind farm's nominal capacity, while the colour represents its type.}
    \label{Fig:farm_map}
\end{figure}

Other measures were taken to increase the observational dataset's reliability. The supplementary material discusses these measures in detail. As a result, some wind farms were completely excluded. Ultimately, \numberoffarms wind farms in Great Britain (34 onshore and 39 offshore) were retained. This study used data from 2019 to 2023. Figure \ref{Fig:farm_map} shows the location, nominal capacity, and type of the wind farms studied.

\subsection{NWP data}
\label{sec:studycase_nwp}

The European Centre for Medium-Range Weather Forecasts (ECMWF) provides the deterministic and ensemble \acrshort{nwp} models used in this work. The deterministic \acrshort{nwp} is from the HRES model, and the ensemble \acrshort{nwp} is from the ENS model. ENS model provides fifty exchangeable ensemble members. HRES was retrieved from the operational archive at full spatial and temporal resolution, but ENS dataset was retrieved from the TIGGE archive, which only stores a limited set of atmospheric variables at 6-hour/0.5$^\circ$ resolution. Retrieving multi-year full-resolution datasets of ENS from the operational archive is prohibitively slow and voluminous. Table \ref{tab:nwp_data} presents the main characteristics of the \acrshort{nwp} models and the atmospheric variables used. The ENS model's time step and computational effort required to perform the forecasts impose restrictions concerning the number of forecast horizons studied in this work. Thus, the forecast horizons studied in this work range from 6 hours to 168 hours ahead by 6-hour frequency.

\begin{table}
\centering
\caption{List of main characteristics of the acquired \acrshort{nwp} datasets, including base times, forecast step frequency, grid resolution, and atmospheric variables used. Note that the wind speed components 100 metres above the surface are unavailable for the ENS model.}
\label{tab:nwp_data}
\scalebox{0.6}{\begin{tabular}{c|c|c} 
\cline{2-3}
\multicolumn{1}{l}{} & \textbf{HRES} & \textbf{ENS} \\ 
\hline
\textbf{Base time} & 00 and 12 UTC & 00 and 12 UTC \\
\rowcolor[rgb]{0.839,0.839,0.839} \textbf{Forecast step frequency} & \begin{tabular}[c]{@{}>{\cellcolor[rgb]{0.839,0.839,0.839}}c@{}}0 to 90 hours ahead by 1\\93 to 144 hours ahead by 3\\150 to 168 hours ahead by 6\end{tabular} & \begin{tabular}[c]{@{}>{\cellcolor[rgb]{0.839,0.839,0.839}}c@{}}6 hours for all forecasting\\horizons\end{tabular} \\
\textbf{Horizontal grid resolution} & 0.1º x 0.1º & 0.5º x 0.5º \\
\rowcolor[rgb]{0.839,0.839,0.839} \textbf{Atmospheric variables} & \begin{tabular}[c]{@{}>{\cellcolor[rgb]{0.839,0.839,0.839}}c@{}}10 metre zonal wind component ($u_{10}$)\\10 metre meridional wind component ($v_{10}$)\\100 metre zonal wind component ($u_{100}$)\\100 metre meridional wind component ($v_{100}$)\\~\end{tabular} & \begin{tabular}[c]{@{}>{\cellcolor[rgb]{0.839,0.839,0.839}}c@{}}10 metre zonal wind component ($u_{10}$)\\10 metre meridional wind component ($v_{10}$)\\2 metre temperature ($T_2$)\end{tabular} \\
\hline
\end{tabular}}
\end{table}

\subsection{Forecast model configurations}
\label{sec:studycase_config}

The method proposed in this work and the reference methods ENS-GBT-None, ENS-QGBT-None and ENS-GBT-EMOS use the same weather-to-power model described in subsection \ref{sec:methodology_a2p}. Therefore, they all use exactly the same \acrshort{nwp} data; see Table \ref{tab:w2p_inputs}. The weather-to-power model is fed by six atmospheric variables from the four grid points closest to the wind farm of interest, totalling 24 inputs.

As discussed in subsection \ref{sec:methodology_a2p}, the ideal scenario would use only \acrshort{nwp} operational analysis and observational data to estimate the weather-to-power model. However, the operational analysis provided by the ensemble \acrshort{nwp} dataset used in this work does not cover all hours of the day to be predicted (i.e., 00, 06, 12, and 18 UTC). Continuous time series were created to overcome this by concatenating the operational analysis and shorter forecast horizon (i.e., 6-hours ahead). Table \ref{tab:w2p_inputs} presents the forecast horizons, base times, and atmospheric variables used to estimate the weather-to-power model.

\definecolor{Alto}{rgb}{0.839,0.839,0.839}
\begin{table}
\centering
\caption{List of base times, forecast horizons, grid points and \acrshort{nwp} variables used for weather-to-power estimation. The \acrshort{nwp} model provides air temperature, and the zonal and meridional components of the wind speed (i.e., $u_{10}$ and $v_{10}$). The variables $w_{10}$, $d_{10}$, and $w_{10}^3$ are estimated from $u_{10}$ and $v_{10}$.}
\label{tab:w2p_inputs}
\scalebox{0.8}{\begin{tblr}{
  row{even} = {Alto},
  column{1} = {c},
  vline{2} = {-}{},
  hline{1,6} = {-}{0.08em},
}
\textbf{NWP base times} & 00 UTC and 12 UTC\\
\textbf{Forecast horizons} & {Operational analysis (i.e., +0 hours ahead) \\ and +6 hours ahead}\\
\textbf{NWP grid points} & 4 closest grid points to the site of interest\\
{\textbf{Direct}\\\textbf{NWP variables}} & {10 metre zonal wind component~$(u_{10})$\\10 metre meridional wind component $(v_{10})$\\2 metre temperature $(T_2)$}\\
{\textbf{Indirect}\\\textbf{NWP variables}} & {10 metre wind speed magnitude, $w_{10} = \sqrt{u_{10}^2 + v_{10}^2}$\\10 metre wind direction, $ d_{10} = \text{arctan2} \left( v_{10}, v_{10} \right)$\\Cube of 10 metre wind speed magnitude, $w_{10}^3$}
\end{tblr}}
\end{table}

The hyperparameters related to the weather-to-power model are tuned for each quantile of interest using the Bayesian Optimization algorithm with random permutation cross-validation \citep{snoek_et_al-2012}. Half of the samples are used for estimation, while the other half is used for testing. Table \ref{tab:w2p_hyperparams} briefly describes the hyperparameters and the search range used in hyperparameter tuning. In the case of hyperparameters not estimated through hyperparameter tuning, the adopted values are presented in Table \ref{tab:w2p_hyperparams}.

\definecolor{Alto}{rgb}{0.839,0.839,0.839}
\begin{table}
\centering
\caption{List of hyperparameters related to the weather-to-power model. The hyperparameter tuning was performed separately for each quantile of interest. Furthermore, they only assume integer values. The hyperparameters associated with the maximum number of features, learning rate, and subsample were not tuned.}
\label{tab:w2p_hyperparams}
\scalebox{0.8}{\begin{tblr}{
  cells = {c},
  row{3} = {Alto},
  row{5} = {Alto},
  row{7} = {Alto},
  vline{2-3} = {-}{0.05em},
  hline{1,9} = {-}{0.08em},
  hline{2} = {-}{0.05em},
}
\textbf{Hyperparameter} & \textbf{Description} & {\textbf{Search range or}\\\textbf{ value adopted}}\\
{Maximum\\  depth} & {Maximum depth, limits \\ the maximum number \\ of nodes in a tree} & {[}5, 9]\\
{Minimum samples\\ to split} & {Minimum number of\\ samples required to\\ split an internal node} & {[}2, 350]\\
{Minimum number\\ of samples at a leaf} & {Minimum number of\\ samples required to \\ be at a leaf node} & {[}2, 350]\\
{Maximum number\\ of features} & {Number of features to\\ consider when looking \\ for the best split} & {Square root of \\ the total number \\ of features}\\
Learning rate & {Controls the contribution\\ of each regression tree in \\ the additive training process} & 0.1\\
{Number of\\ estimators} & {Number of boosting iterations.\\ That is, the definition of base \\ learners in the final ensemble} & {[}2, 150]\\
{Subsample\\ (or bag fraction)} & {Fraction of samples to be\\ used for fitting the individual \\ base learners} & {80\% of the data is \\ used for fitting the \\ individual trees}
\end{tblr}}
\end{table}

The data samples used to estimate the methods based on the weather-to-power model are split to guarantee independence between the data used for estimating the weather-to-power modelling and ensemble combination. The first half of the samples available for model estimation is used to estimate the weather-to-power model (including the hyperparameter tuning). The second half is used to estimate the ensemble combination parameters.

The reference method HRES-QGBT-None, based on \citet{andrade_and_bessa-2017}, uses as input the zonal and meridional components of the wind speed provided by the deterministic \acrshort{nwp} model at two vertical levels, i.e., 10m and 100m above the surface. Its horizontal domain comprises 5x5 \acrshort{nwp} grid points centred on the site of interest. The number of \acrshort{nwp} grid points used here is lower than that used by \citet{andrade_and_bessa-2017}. However, the area comprised by the grid points remains the same due to the higher spatial resolution of the \acrshort{nwp} used in \citet{andrade_and_bessa-2017}. Hyperparameter tuning was performed separately for each forecast horizon and quantile. Table \ref{tab:HRES-QGBT-None_hyperparams} presents the search range used in hyperparameter tuning. Table \ref{tab:HRES-QGBT-None_hyperparams} presents the adopted values in the case of hyperparameters not estimated through hyperparameter tuning.

\definecolor{Alto}{rgb}{0.839,0.839,0.839}
\begin{table}
\centering
\caption{List of hyperparameters related to the reference method HRES-QGBT-None. The hyperparameter tuning was performed separately for each quantile and forecast horizon. Furthermore, they only assume integer values. The hyperparameters associated with the maximum number of features, learning rate, and subsample were not tuned.}
\label{tab:HRES-QGBT-None_hyperparams}
\scalebox{0.8}{\begin{tblr}{
  cells = {c},
  row{3} = {Alto},
  row{5} = {Alto},
  row{7} = {Alto},
  vline{2} = {-}{},
  hline{1,9} = {-}{0.08em},
  hline{2} = {1}{},
  hline{2} = {2}{r},
}
\textbf{Hyperparameter} & {\textbf{Search range or}\\\textbf{ value adopted}}\\
{Maximum\\ depth} & {[}5, 9]\\
{Minimum samples\\ to split} & {[}10, 160]\\
{Minimum number \\ of samples at a leaf} & {[}10,110]\\
{Maximum number\\ of features} & {Square root of the total\\ number of features}\\
Learning rate & 0.1\\
{Number of\\ estimators} & {[}50, 400]\\
subsample & {80\% of the data is used for \\ fitting the individual trees}
\end{tblr}}
\end{table}

Due to the disparity between the characteristics of the ENS and HRES datasets, it is challenging to compare HRES-QGBT-None and the other studied methods. Thus, a modification of HRES-QGBT-None is included subject to constraints similar to those imposed by the TIGGE archive on the ENS dataset. This constrained version of HRES-QGBT-None is called HRESc-QGBT-None. The modifications include using atmospheric variables at only one vertical level (i.e., 10 meters above the surface) and forecasting horizons with a 6-hour step frequency. Note that the horizontal grid resolution was not changed.

All \acrshort{gbt}s used in this work were estimated using the implementation provided by \citet{scikit-learn}, and the Bayesian Optimization Algorithm used for hyperparameter tuning was provided by \citet{skopt}.

\subsection{Case study periods}
\label{sec:studycase_periods}

In this work, time series are divided into two distinct sets: estimation and test. The estimation period is used to estimate the weather-to-power and ensemble calibration/combination models. In contrast, the test period is used exclusively to assess forecasting performance. No estimation or decision-making is made during this period, ensuring a clear evaluation of the forecasting models. The estimation period always comprises a time window before the testing period. 

To maximise the utility of available data, we repeat the study for three distinct test periods: 2021, 2022, and 2023 with different estimation periods from 2 to 4 years, depending on data availability. The case studies are named according to the test period and estimation window. For instance, ``2023-4y’’ refers to 2023 as the test period and the four preceding years (i.e., 2019, 2020, 2021, and 2022) as the estimation period.

\FloatBarrier
\section{Results and discussion}
\label{sec:results}

The results and discussion are divided into three subsections. The first presents the results of the reference methods based on deterministic and ensemble \acrshort{nwp} models and discusses the added value of using ensemble \acrshort{nwp} and the relevance of different uncertainty sources across different forecast horizons. The second presents the results related to the proposed method and compares its performance with the reference methods. The third subsection discusses forecast performance under high \acrshort{nwp} uncertainty.

\subsection{Performance of the reference methods}
\label{sec:results_ref_performance}

\begin{figure}[h]
    \centering
    \subcaptionbox{Berry Burn Windfarm (BMU ID: E\_BRYBW-1). Onshore farm consisting of 29 wind turbine generators, totalling a nominal capacity of 66.7 MW.}
    {\includegraphics[scale=0.57, trim={0 0 0 0}, clip]{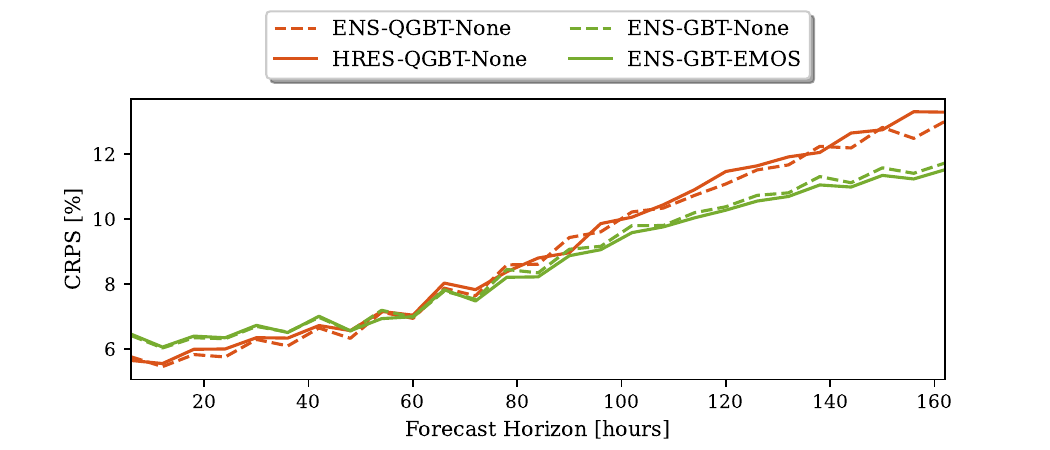}}
   
    \subcaptionbox{London Array Windfarm Unit 4 (BMU ID: T\_LARYW-4). Offshore farm consisting of 50 wind turbine generators, totalling a nominal capacity of 180 MW.}
    {\includegraphics[scale=0.57, trim={0 0 0 1.5cm}, clip]{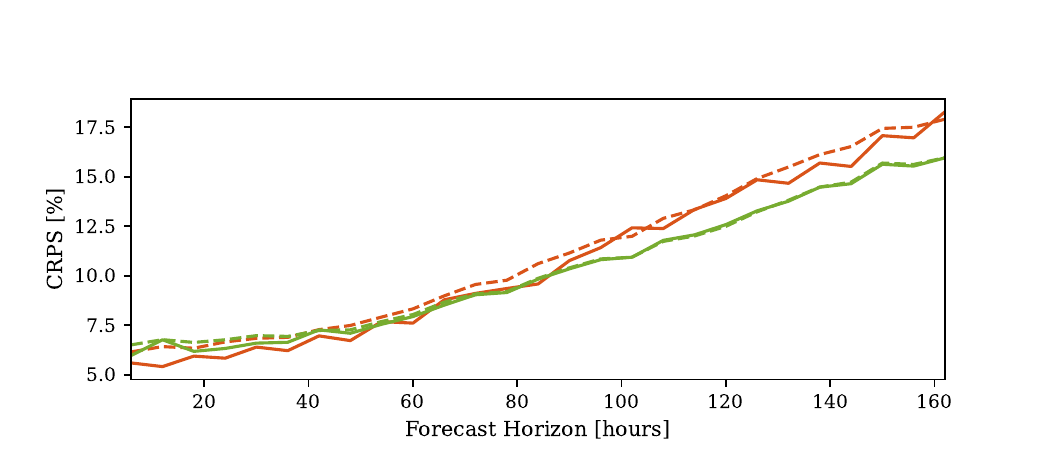}}

    \caption{CRPS of the reference methods over the forecast horizons. The results presented concern the case study 2021-2y. The short-term methods are plotted in red, and the mid-term methods are in green.}
    \label{Fig:ex_farm_classification}
\end{figure}

This work adopts four reference forecast methods introduced in Section \ref{sec:methodology_refmodels}; three based on the ensemble \acrshort{nwp}, and one based on the deterministic \acrshort{nwp}. The reference methods can be grouped according to how their performance degrades with increasing forecast horizons.
The first group have relatively good performance for short horizons but suffers rapid degradation for longer horizons, whereas the second group perform less well for short horizons but does not degrade as much for longer horizons.
Therefore, the first group, comprising HRES-QGBT-None and ENS-QGBT-None, is referred to as short-term methods, and the second comprising ENS-GBT-EMOS and ENS-GBT-None, is referred to as mid-term.
Figure \ref{Fig:ex_farm_classification} illustrates the performance of the two groups of reference methods over the forecast horizons. 

\begin{figure}[h]
    \centering
    \subcaptionbox{CRPS skill score of HRES-QBGT-None with respect to the ENS-QBGT-None considering the case studies 2021-2y, 2022-3y and 2023-4y.}
    {\includegraphics[scale=0.57, trim={0 0 0 0.7cm}, clip]{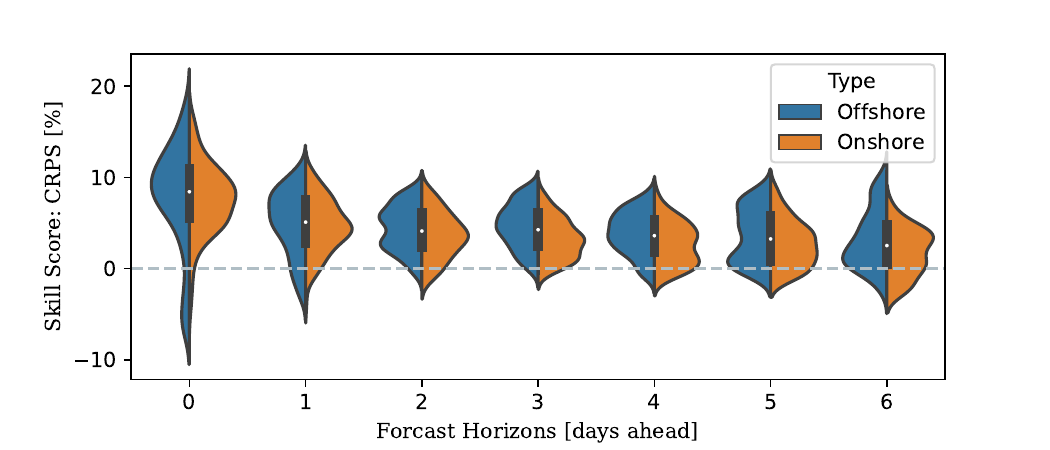}}
   
    \subcaptionbox{CRPS skill score of HRESc-QGBT-None with respect to the ENS-QBGT-None considering the case study 2021-2y.}
    {\includegraphics[scale=0.57, trim={0 0 0 0.7cm}, clip]{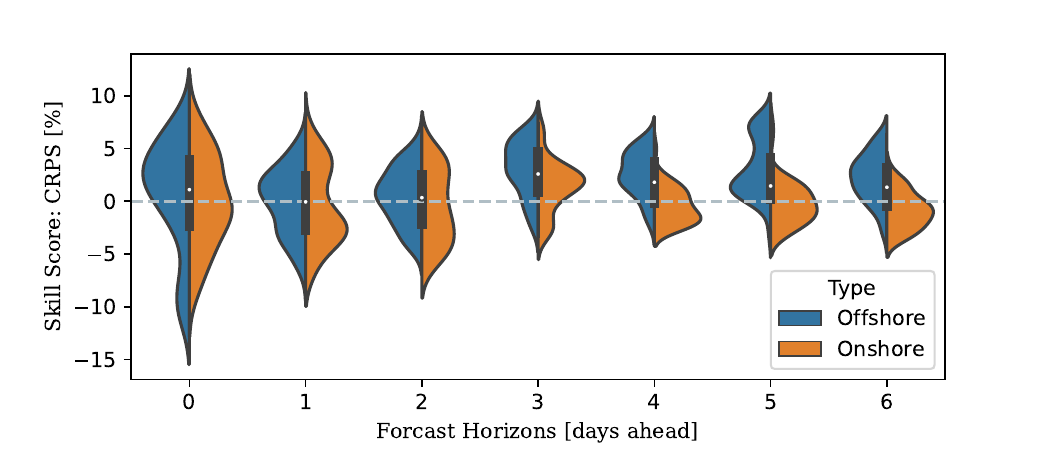}}

    \caption{CRPS skill score comparison between the short-term reference methods. Each forecast day considers the average skill score of all included forecast horizons. Figure (b) only presents the results related to the 2021-2y case study. Due to the high computational effort required, the reference method HRESc-QGBT-None was estimated solely for this case study.}
    \label{Fig:SS_short-term}
\end{figure}

%
Of the short-term methods, HRES-QGBT-None outperforms ENS-QGBT-None for all forecast horizons and case studies, shown in Figure \ref{Fig:SS_short-term}a. The better performance is likely due to the high temporal resolution and the larger set of atmospheric variables available from HRES compared to ENS (from TIGGE). The larger information set provides a more accurate description of the wind close to the turbine hub height, and the description of wind speed at two vertical levels enables better discrimination between atmospheric conditions by the weather-to-power model (i.e., QGBT). Moreover, the HRES model's high temporal resolution enables the extraction of temporal features that have proven effective for achieving high performance, especially in shorter forecast horizons \cite{andrade_and_bessa-2017}. 

Figure \ref{Fig:SS_short-term}(b) shows that restricting HRES-QGBT-None to the same resolution and atmospheric variables as ENS-QGBT-None, denoted HRESc-QGBT-None, leads to a similar outcome for offshore wind farms, but only some onshore wind farms. Features engineered from the full-resolution deterministic \acrshort{nwp} are more predictive than those from reduced-resolution, suggesting that all ENS-based methods presented here could be improved with access to the full-resolution model data.
Due to the superior performance of HRES-QBGT-None, it will be considered the short-term forecasting reference method.

\begin{figure}[hbt!]
    \centering
    \includegraphics[scale=0.57, trim={0 0 0 0.7cm}, clip]{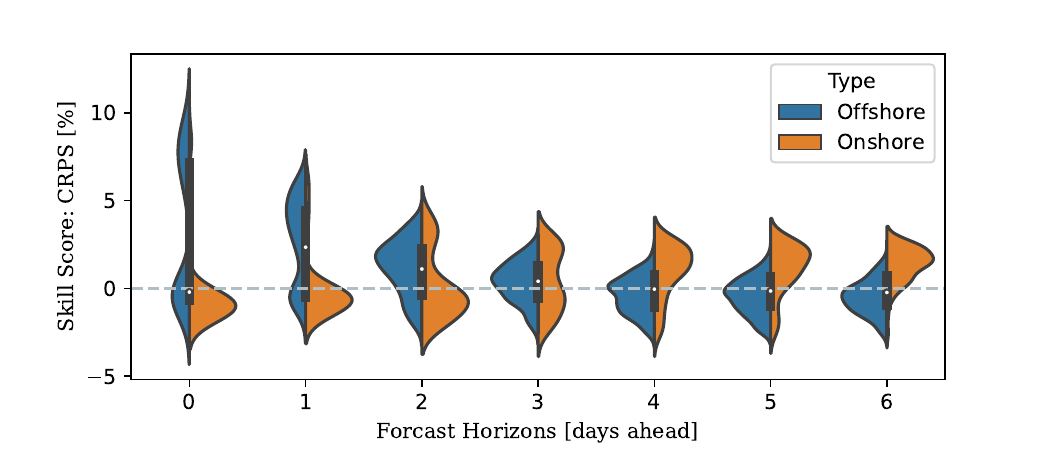}
    \caption{CRPS skill score of ENS-GBT-EMOS with respect to the ENS-GBT-None for the case studies 2021-2y, 2022-3y and 2023-4y. Each forecast day considers the average skill score of all included forecast horizons.}
    \label{Fig:SS_mid-term}
\end{figure}

The comparison between the mid-term methods can be interpreted as performance improvement due to \acrshort{emos} since both methods differ in this aspect. Figure \ref{Fig:SS_mid-term} shows that the performance depends on the wind farm type. ENS-GBT-EMOS outperforms ENS-GBT-None at up to three-day-ahead horizons for offshore wind farms while narrowly underperforming for onshore wind farms. The bimodal behaviour of offshore farms in the first two days is due to a group of wind farms that benefited more from the \acrshort{emos} correction. The same occurs with the onshore farms two and three days ahead. For horizons from 3 days ahead, ENS-GBT-EMOS outperforms ENS-GBT-None for onshore wind farms and obtains similar results for offshore ones. As discussed later, the weather-to-power uncertainty is more relevant than the \acrshort{nwp} model uncertainty for short forecast horizons. However, the mid-term reference methods do not consider the weather-to-power uncertainty in the modelling process. Thus, the performance improvement due to \acrshort{emos} calibration in the first horizons is mainly attributed to correcting the omission of weather-to-power uncertainty during the modelling process. On the other hand, the performance improvement in the mid-term horizons is mainly due to the correction of the ensemble spread, which improves the model's reliability. Due to the overall performance presented by ENS-GBT-EMOS, it will now be considered the mid-term forecasting reference method.

\begin{figure}[hbt!]
    \centering
    \includegraphics[scale=0.57, trim={0 0 0 0.7cm}, clip]{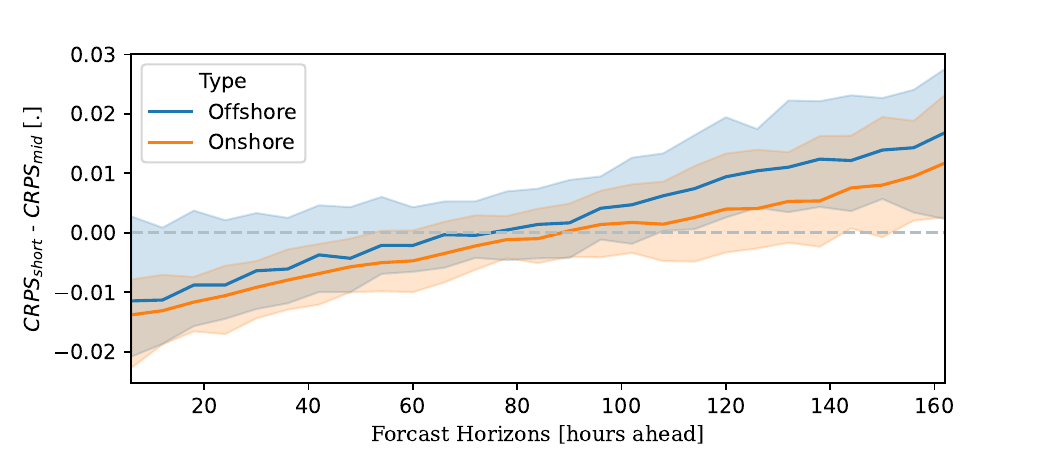}
    \caption{CRPS difference between short- and mid-term reference methods considering the 2021-2y, 2022-3y and 2023-4y case studies. Each line represents the median value, and the filled area represents the interval between the 5th and 95th percentiles.}
    \label{Fig:crps-diff_short-mid}
\end{figure}

Mid-term reference methods perform better at longer forecast horizons across all wind farms and case studies. However, the horizon at which mid-term methods outperform short-term methods varies depending on the wind farm. Thus, this study examines the difference in \acrshort{crps} between short- and mid-term methods over the horizons to determine when the ensemble \acrshort{nwp} benefits forecast performance.
%
For onshore wind farms, the mid-term method outperforms the short-term method from around four-days ahead, whereas for offshore wind farms, the mid-term method tends to outperform the short-term method from three days-ahead, and for some offshore wind farms, the mid-term method performs best for all horizons (see Figure \ref{Fig:crps-diff_short-mid}).

\begin{figure}[t!]
    \centering
    \subcaptionbox{Berry Burn Windfarm (BMU ID: E\_BRYBW-1). Onshore farm consisting of 29 wind turbine generators, totalling a nominal capacity of 66.7 MW.}
    {\includegraphics[scale=0.57, trim={0 0 0 0}, clip]{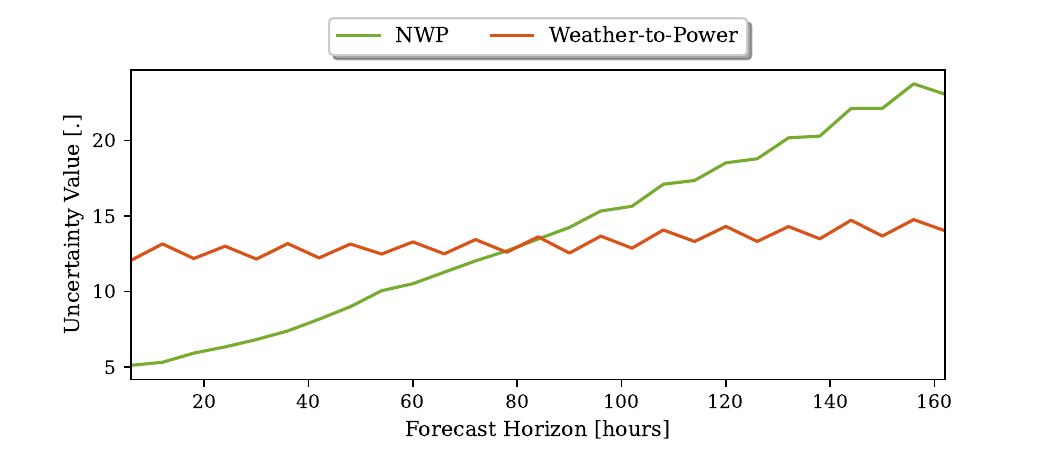}}
   
    \subcaptionbox{London Array Windfarm Unit 4 (BMU ID: T\_LARYW-4). Offshore farm consisting of 50 wind turbine generators, totalling a nominal capacity of 180 MW.}
    {\includegraphics[scale=0.57, trim={0 0 0 1.1cm}, clip]{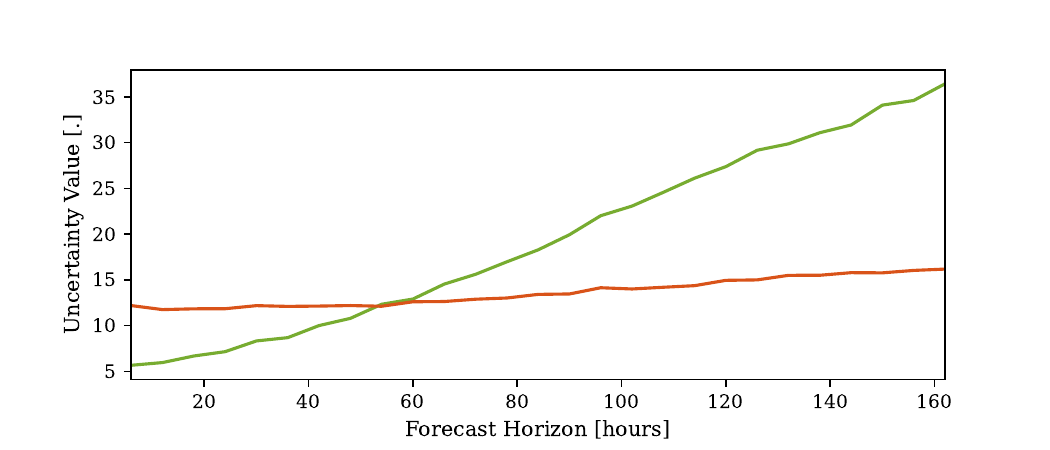}}

    \caption{Uncertainty value over the forecast horizons. The results presented concern the case study 2021-2y. Note that the case study and wind farms presented here are the same as those in Figure \ref{Fig:ex_farm_classification}.}
    \label{Fig:ex_uncertainty}
\end{figure}

The performance difference between short- and mid-term methods is closely related to the sources of uncertainty considered in the modelling process. Figure \ref{Fig:ex_uncertainty} exemplifies the relationship between forecast horizons and uncertainty values. For more information on uncertainty value, see subsection \ref{sec:assessment_framework}. \acrshort{nwp} uncertainty increases with increasing forecast horizons. This increase is due to the inherently chaotic nature of the atmospheric system, which is well documented in the literature \citep{buizza_et_al-1999}. On the other hand, weather-to-power uncertainty presents small variability over the forecast horizons. The small variability is because weather-to-power uncertainty is affected by diurnal patterns in weather conditions. 

Weather-to-power uncertainty tends to be greater than \acrshort{nwp} uncertainty in shorter horizons, i.e., up to 48 hours ahead. Thus, forecasting methods that do not consider weather-to-power uncertainty during the modelling process have difficulty reliably forecasting these horizons. On the other hand, \acrshort{nwp} uncertainty is greater for longer horizons, meaning that additional information provided by ensemble \acrshort{nwp} has a greater impact for these horizons. 

\begin{figure}[hbt!]
    \centering
    \includegraphics[scale=0.57, trim={0 0 0 0.7cm}, clip]{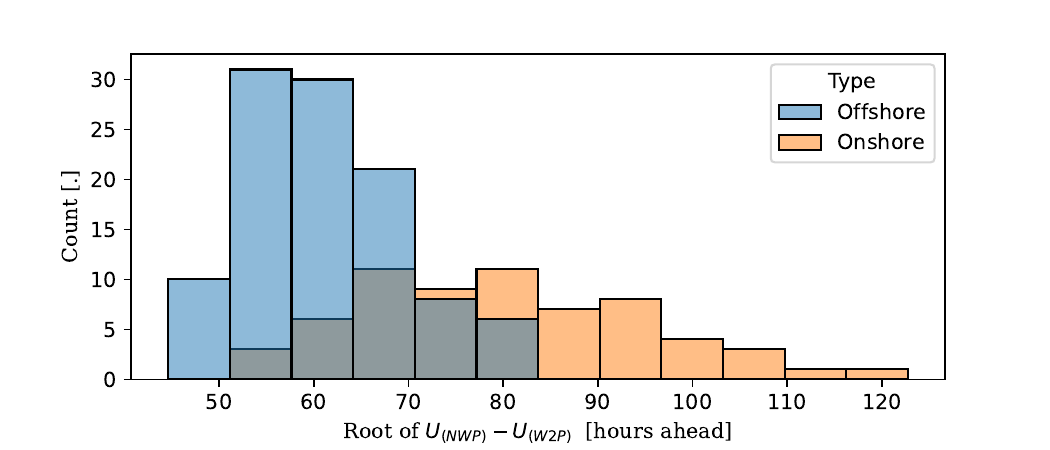}
    \caption{Histogram of the horizon at which $U^{(NWP)}=U^{(W2P)}$ by wind farm.
    The results present in this figure combine the 2021-2y, 2022-3y and 2023-4y case studies.}
    \label{Fig:poly_root_uncertainty}
\end{figure}

The difference between $U^{(NWP)}$ and $U^{(W2P)}$ over the forecast horizons shows dependency concerns on the farm type. However, due to the seasonality presented by $U^{(W2P)}$, it becomes difficult to analyse which horizon $U^{(NWP)}$ becomes greater than $U^{(W2P)}$. Thus, Figure \ref{Fig:poly_root_uncertainty} shows the root of the first-degree polynomial adjusted to the difference curve between $U^{(NWP)}$ and $U^{(W2P)}$. The dominance of \acrshort{nwp} uncertainty tends to occur more frequently at shorter horizons for offshore farms than onshore ones. Thus, the relationship between the uncertainty difference and the farm type aligns with the performance difference. Therefore, the findings in Figure \ref{Fig:poly_root_uncertainty} support the dependence hypothesis between the forecasting performance and the uncertainty sources considered in the modelling process.

Even though the performance difference between short- and mid-term methods is linked to the type of wind farm, it shows significant variability. Thus, a seamless forecasting method capable of achieving state-of-the-art performance across all desired forecast horizons is highly valuable. However, achieving consistently state-of-the-art performance across a wide range of forecast horizons using just one predictive method is challenging as uncertainties associated with both \acrshort{nwp} and weather-to-power must be handled effectively.

\FloatBarrier
\subsection{Performance of the proposed method}
\label{sec:results_probw2p_performance}

This study introduces a novel method to explicitly quantify and combine uncertainties related to \acrshort{nwp} and weather-to-power. It aims to achieve state-of-the-art performance for short- and mid-term horizons (i.e., from 6 to 162 hours ahead). State-of-the-art performance is given by the reference method that achieves the best performance for a given horizon, with HRES-QBGT-None and ENS-GBT-EMOS representing short- and mid-term, respectively. The best reference method is independently chosen for each case study, wind farm, and forecast horizon based on their performance during the test period. Note that this is based on the test period results and consequently, it represents a theoretical reference that is conservative in favour of the reference methods. 

\begin{figure}[h]
    \centering

    \subcaptionbox{CRPS skill score of the proposed method with respect to the HRESc-QGBT-None.}
    {\includegraphics[scale=0.57, trim={0 0 0 0}, clip]{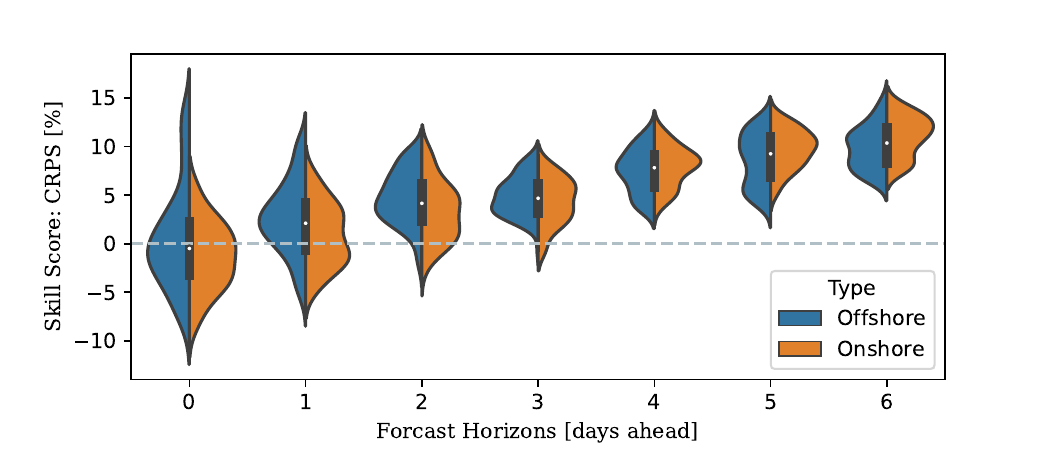}}
   
    \subcaptionbox{CRPS skill score of the proposed method with respect to the ENS-GBT-EMOS.}
    {\includegraphics[scale=0.57, trim={0 0 0 0}, clip]{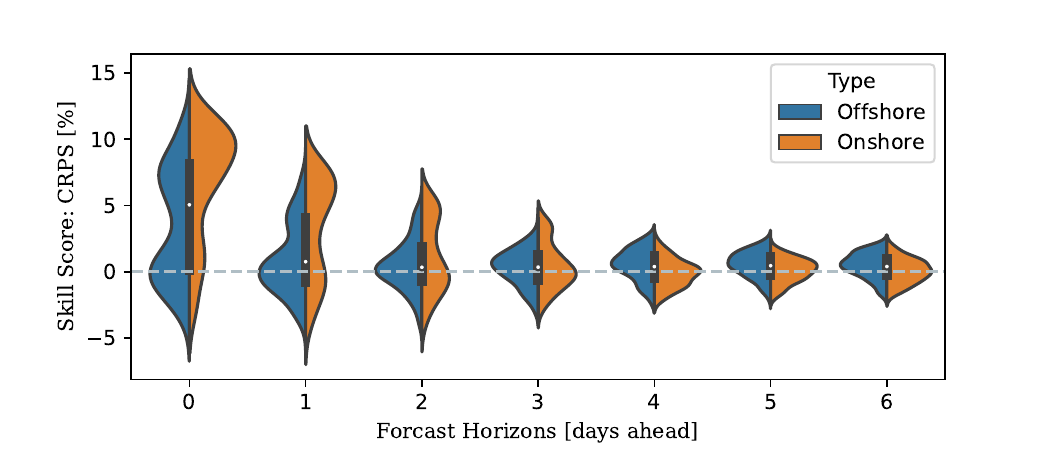}}

    \subcaptionbox{CRPS skill score of the proposed method with respect to the state-of-the-art. The state-of-the-art is represented by the reference methods HRESc-QGBT-None and ENS-GBT-EMOS.}
    {\includegraphics[scale=0.57, trim={0 0 0 0}, clip]{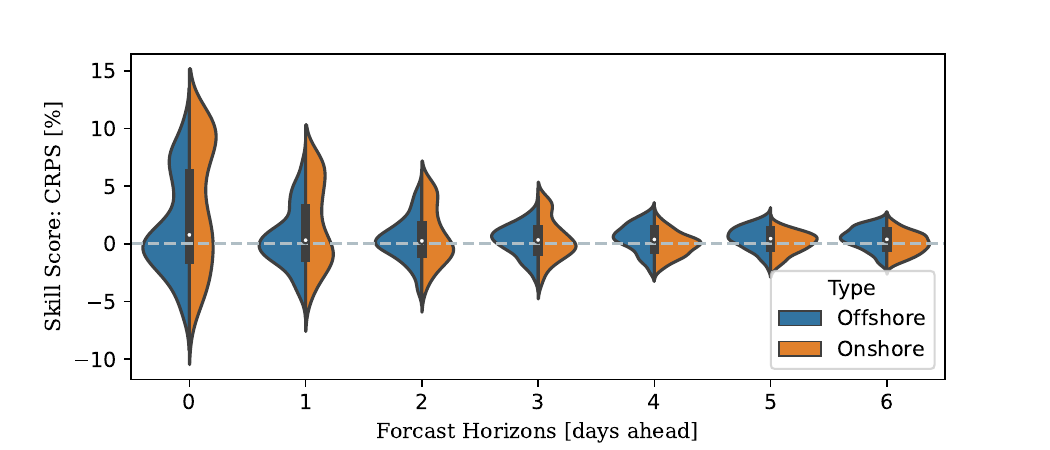}}
    
    \caption{CRPS skill score of the proposed method with respect to the short-term method, mid-term method, and state-of-the-art. The ensemble and the deterministic \acrshort{nwp} models have the same temporal resolution and set of atmospheric variables. Note that the deterministic \acrshort{nwp} model still has a high spatial resolution. Each forecast day considers the skill score of all included forecast horizons. This Figure presents the results related to the case study 2021-2y. Due to the high computational effort required, the reference method HRESc-QGBT-None was estimated solely for this case study.}
    \label{Fig:ss_final_cmp}
\end{figure}

Figure \ref{Fig:ss_final_cmp} presents the proposed method's \acrshort{crps} skill score concerning the short-term method, mid-term method, and state-of-the-art when using deterministic and ensemble \acrshort{nwp} models with the same temporal resolution and set of atmospheric variables.

The proposed method performs similarly to the short-term reference method on the first day and outperforms it from two days-ahead. The improvement occurs for horizons in which \acrshort{nwp} uncertainty becomes relevant. On the other hand, the proposed method outperforms the mid-term method in horizons with relevant weather-to-power uncertainty, i.e., up to one day-ahead. Consequently, the proposed method achieved state-of-the-art performance along all studied horizons.

\begin{figure}[h]
    \centering

    \subcaptionbox{CRPS skill score of the proposed method with respect to the HRES-QGBT-None.}
    {\includegraphics[scale=0.57, trim={0 0 0 0}, clip]{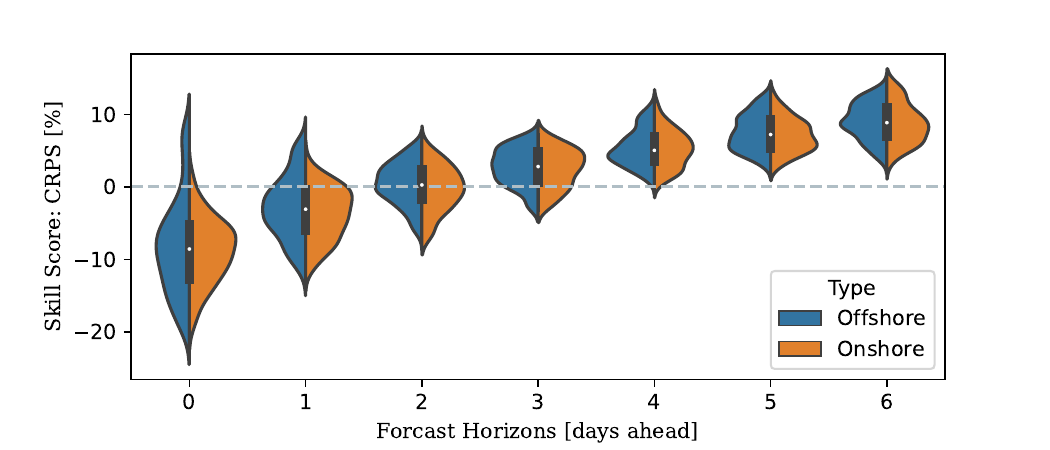}}

    \subcaptionbox{CRPS skill score of the proposed method with respect to the state-of-the-art. The state-of-the-art is represented by the reference methods HRES-QGBT-None and ENS-GBT-EMOS.}
    {\includegraphics[scale=0.57, trim={0 0 0 0}, clip]{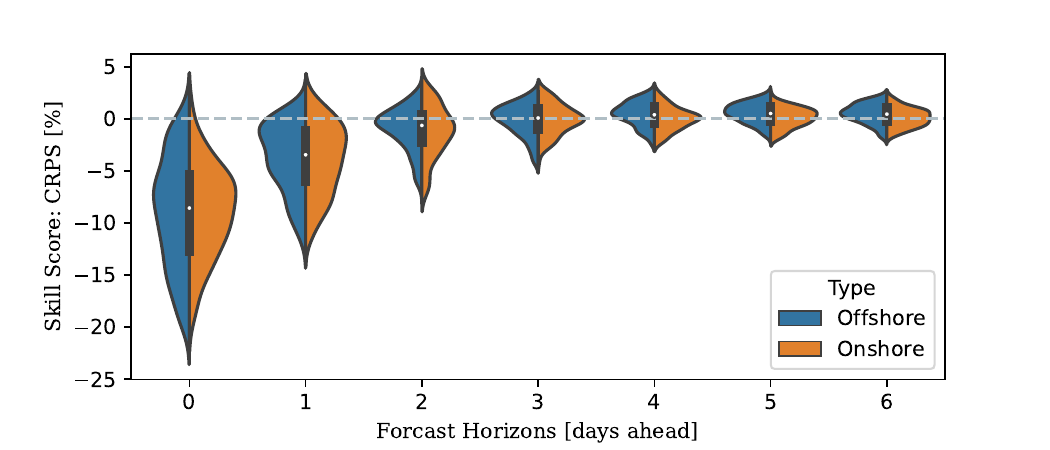}}
    
    \caption{CRPS skill score of the proposed method with respect to the short-term method and state-of-the-art. The state-of-the-art is represented by the reference methods HRES-QGBT-None and ENS-GBT-EMOS. Each forecast day considers the average skill score of all included forecast horizons. This Figure presents the results of the case studies 2021-2y, 2022-3y and 2023-4y.}
    \label{Fig:ss_final_highres}
\end{figure}

However, the short-term reference method significantly improves accuracy when using a deterministic \acrshort{nwp} model with higher temporal resolution and a larger set of atmospheric variables (see Figure \ref{Fig:ss_final_highres}). This improvement occurs because, as previously discussed, the high temporal resolution enables the extraction of temporal features that have proven effective in achieving high performance in horizons up to 48 hours ahead. Additionally, the larger set of atmospheric variables improves the classification performed by the \acrshort{gbt}, leading to performance enhancement. Consequently, the short-term method is better than the proposed method for horizons up to the day ahead. However, despite high spatial and temporal resolution, the short-term method cannot outperform the proposed method for horizons beyond one day ahead, where weather-to-power uncertainty is relevant.

The results shown in Figure \ref{Fig:ss_final_highres} indicate that the accuracy of the proposed method could be significantly enhanced by utilising the full-resolution ensemble \acrshort{nwp} and a wider range of atmospheric variables, given that both the weather-to-power model employed by HRES-QGBT-None and the proposed method are based on \acrshort{gbt}s. Therefore, the authors suggest investigating the performance improvement using high-resolution ensemble \acrshort{nwp}.

The proposed method places great emphasis on the necessary computational effort. Among all the models studied in this work, the \acrshort{gbt} used to model the weather-to-power relationship requires the most computational effort to estimate. This is attributed to the estimation of multiple \acrshort{gbt}s and hyperparameter tuning.

The ENS-GBT-EMOS and ENS-GBT-None reference methods only require the estimation of a single \acrshort{gbt}, regardless of the number of quantiles or forecast horizons. ENS-QGBT-None requires the estimation of one \acrshort{gbt} for each quantile of interest, regardless of the number of forecast horizons. On the other hand, HRES-QGBT-None requires the estimation of one \acrshort{gbt} for each quantile of interest and each forecasting horizon. However, the proposed method requires the estimation of only three \acrshort{gbt}s, regardless of the quantiles of interest and forecast horizons (see Table \ref{tab:com_effort}). As a result, the proposed method achieves state-of-the-art performance while requiring low computational effort, especially when compared to HRES-QGBT-None.

\begin{table}[]
\centering
\caption{Number of GBTs estimated for each method per wind farm for 19 quantiles of interest and 27 forecast horizons.}
\label{tab:com_effort}
\begin{tabular}{c|c}
\hline
\multicolumn{1}{l|}{} & \multicolumn{1}{l}{\textbf{Number of \acrshort{gbt}s}} \\ \hline
\rowcolor[HTML]{D5D5D5} 
ENS-GBT-None & 1 \\
ENS-QGBT-None & 19 \\
\rowcolor[HTML]{D5D5D5} 
HRES-QGBT-None & 513 \\
ENS-GBT-EMOS & 1 \\
\rowcolor[HTML]{D5D5D5} 
\textbf{ENS-QGBT-$\beta$MM} & \textbf{3} \\ \hline
\end{tabular}
\end{table}

\subsection{Forecasting under variable NWP uncertainty}
\label{sec:results_probw2p_high_unc}

\begin{figure}[hbt!]
    \centering

    \subcaptionbox{Magnitude of the wind speed predicted by the ensemble (ENS) and deterministic (HRESc) NWP models}
    {\includegraphics[scale=0.57, trim={0 0 0 0}, clip]{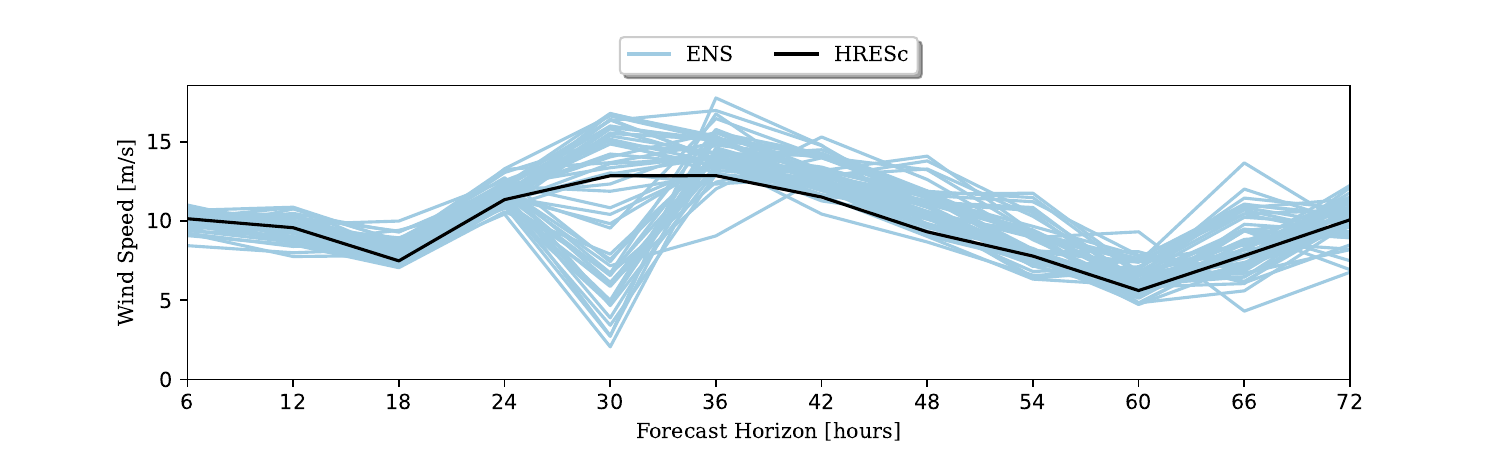}}

    \subcaptionbox{Intervals predicted by ENS-QGBT-$\beta$MM}
    {\includegraphics[scale=0.57, trim={0 0 0 0}, clip]{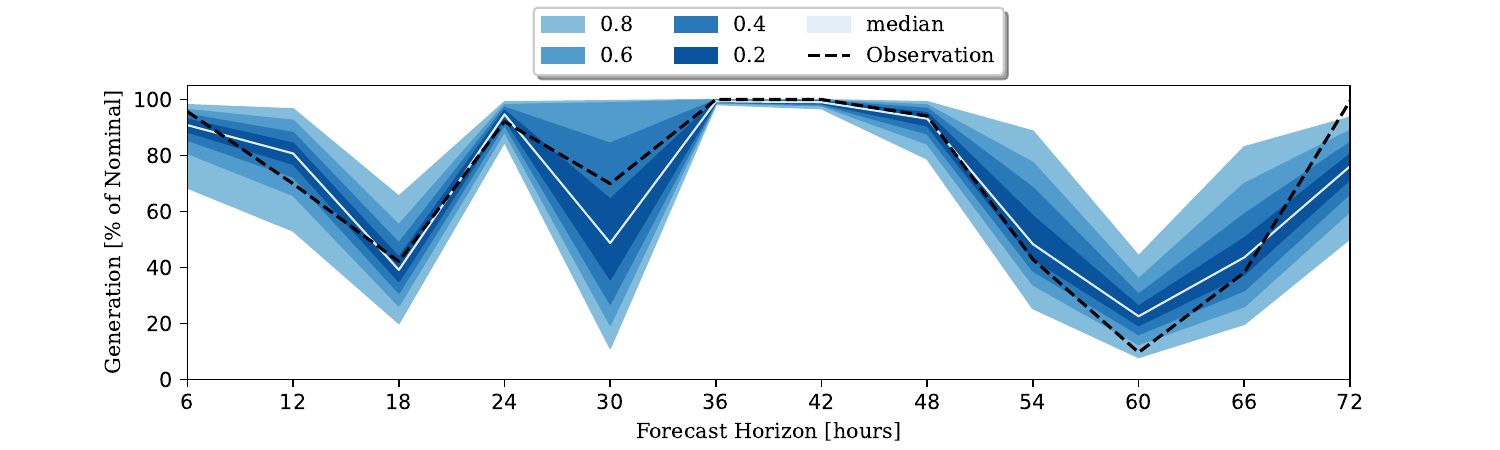}}

    \subcaptionbox{Intervals predicted by HRESc-QGBT-None.}
    {\includegraphics[scale=0.57, trim={0 0 0 1.4cm}, clip]{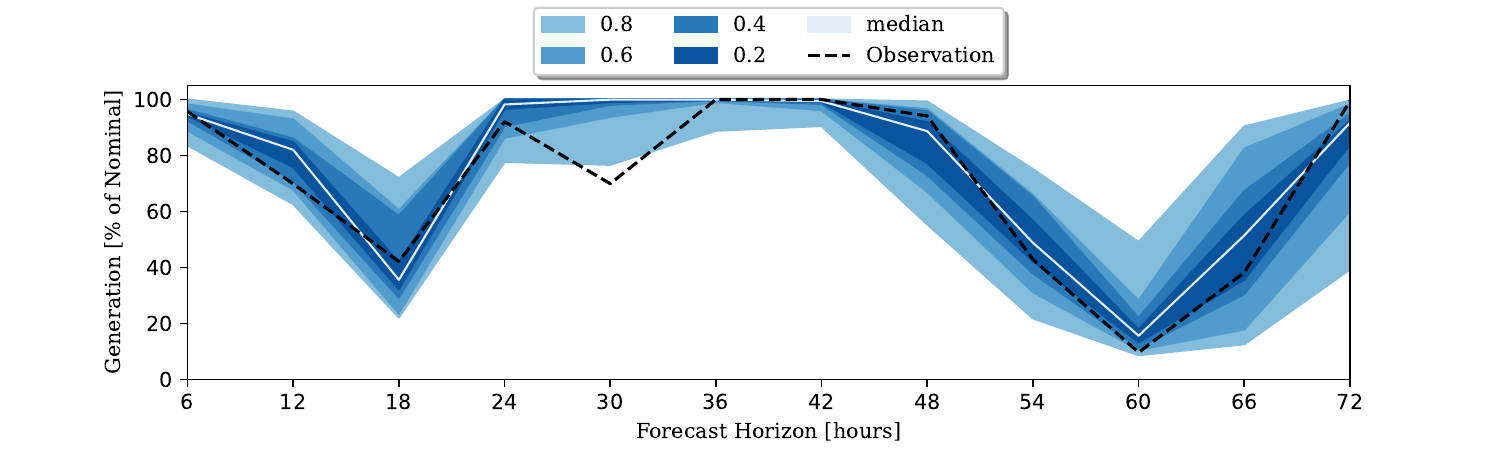}}
    
    \caption{Example of high NWP uncertainty related to the wind farm Mocambre Offshore unit 2 (BMU ID: T\_GRGBW-3) for the NWP base time 2021-12-09 00:00 and case study 2021-2y. The legend explains the colours assigned to the prediction intervals.}
    \label{Fig:example_high_unc}
\end{figure}

The results presented and discussed so far originate from an overall performance assessment, as they represent average performance over multiple years. However, some events are of greater interest to decision-makers. In particular, forecasts subject to high \acrshort{nwp} uncertainty are important. As discussed previously, \acrshort{nwp} uncertainty tends to increase with forecast horizons due to the atmospheric system's inherently chaotic nature. However, some atmospheric conditions are associated with high wind power forecast uncertainty, even for short horizons, such as the arrival time or precise location of a frontal system. Ensemble \acrshort{nwp} may reflect this uncertainty through the spread of ensemble members, while deterministic \acrshort{nwp} provides only a single weather trajectory.
Figure \ref{Fig:example_high_unc} presents an example of high \acrshort{nwp} uncertainty occurring at the relatively short lead-time of 30 hours ahead which is captured by ensemble but not deterministic \acrshort{nwp}.

\begin{figure}[h]
    \centering
    \includegraphics[scale=0.57, trim={0 0 0 0}, clip]{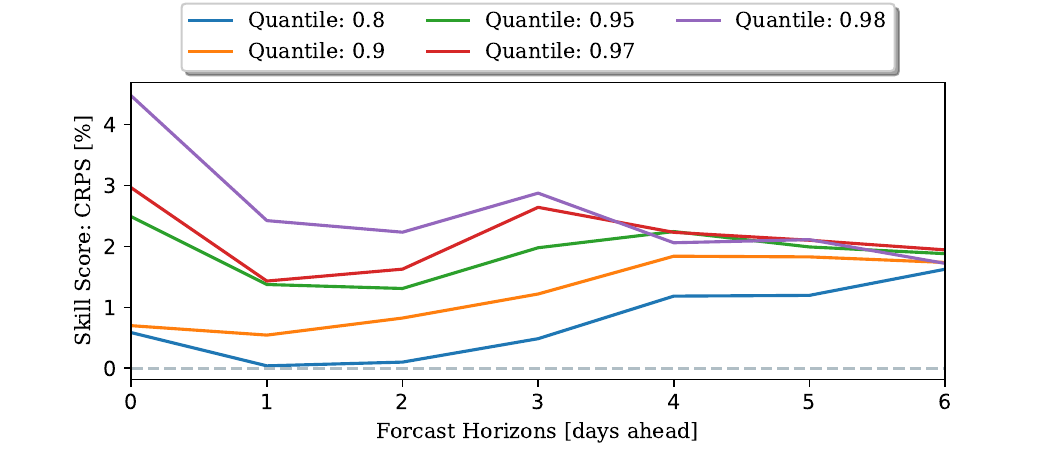}

    \caption{CRPS skill score of the proposed method with respect to the state-of-the-art under different scenarios of NWP uncertainty. The state-of-the-art is composed of the reference methods HRESc-QGBT-None and ENS-GBT-EMOS. This figure shows the results related to the case studies 2021-2y. Each scenario concerns the timestamps in which the NWP uncertainty was greater than a threshold given by a quantile. The quantile is estimated distinctly for each wind farm, case study, and forecast horizon. Each forecast day considers the skill score of all included forecast horizons. Each line represents the median of the skill score.}
    \label{Fig:ss_high_nwp_unc}
\end{figure}

This study analyses five different scenarios. Each concerns periods in which the \acrshort{nwp} uncertainty was greater than a given quantile of past uncertainty (i.e., inter-quartile range of wind power ensemble, see Eq. \ref{eq:unc_nwp}), calculated for each wind farm, case study, and forecast horizon. The proposed and reference methods during these periods are compared in Figure \ref{Fig:ss_high_nwp_unc}. When using a deterministic \acrshort{nwp} model with the same temporal resolution and set of atmospheric variables, the proposed method outperforms the state-of-the-art at all time horizons (Figure \ref{Fig:ss_high_nwp_unc}). For zero and one days-ahead, an improvement compared to the state-of-the-art is present, which was not evident in the long-run average \acrshort{crps}. Because the short-term reference method does not consider the \acrshort{nwp} uncertainty, it is not able to discriminate between high/low weather uncertainty and performance sufferers as a result.

While infrequent, such events have a disproportionately large impact on decision-makers who can only take mitigating action if supplied with an informative forecast. Therefore, although it is still not considered, assessing forecast performance under high-uncertainty events should be considered in recommended practices such as \citet{iea_selection}. This highlights the significant value added by ensemble \acrshort{nwp} that is not necessarily reflected in conventional metrics.



\FloatBarrier
\section{Conclusion}
\label{sec:conclusion}

Wind power forecasts are subject to uncertainty originating from weather forecasts and weather-to-power conversion. This work has shown how weather-to-power uncertainty dominates short-term forecast performance, while weather forecast uncertainty dominates mid-term. Typically, the transition from one situation to the other is two- to three-days ahead, but can vary dramatically between wind farms. The transition typically occurs at shorter lead-times for offshore wind farms compared to onshore.
    
The method proposed here explicitly quantifies both sources of uncertainty and achieves state-of-the-art forecast performance across all horizons, where previously separate models/methods were necessary. Furthermore, performance during periods of high weather uncertainty, such as frontal passages, is improved in short-term forecasts, which state-of-the-art methods based on deterministic \acrshort{nwp} fail to capture.

Results are based on an extensive dataset comprising 73 onshore and offshore wind farms in Great Britain over five years, i.e., from 2019 to 2023. The numerical weather forecasts are from deterministic and ensemble operational \acrshort{nwp} models provided by the ECMWF, targeting horizons from 6 to 162 hours ahead.

The proposed method employs non-parametric uncertainty quantification for weather-to-power based on gradient-boosted quantile regression trees, which control the shape of kernels used to convert ensemble numerical weather prediction into non-parametric density forecasts of wind power production. Kernels are combined using a beta-transformed linear opinion pool to address ensemble mis-calibration and potential dependence between kernels.

Despite the encouraging results presented in this paper, the forecast method proposed here could be enhanced in several ways. First, through using a high-resolution ensemble \acrshort{nwp} model to improve the feature engineering. Additionally, modelling the dependency between kernels instead of using opinion pools could increase the performance by allowing for non-parametric posterior distributions. Finally, performing member-by-member correction to retain spatio-temporal structure in ensemble members would allow for spatio-temporal coherence between forecasts from different wind farms.

\section*{Acknowledgements}

Gabriel Dantas is supported by a PhD scholarship from the University of Glasgow School of Mathematics and Statistics. The authors would like to express their gratitude to the Balancing Mechanism Reporting Agent and the European Centre for Medium-Range Weather Forecasts for providing the data used in this work.

For the purpose of open access, the authors have applied a Creative Commons Attribution (CC BY) licence to any Author Accepted Manuscript version arising from this submission.

\section*{Data availability}
\label{sec:data_availability}

All datasets, codes and results are available at Zenodo. Table \ref{tab:repositories} lists the repositories and their content. Contains BMRS data \copyright Elexon Limited copyright and database right 2024. ECMWF data is licenced under Creative Commons Attribution 4.0 International (CC BY 4.0) and is subject to ECMWF terms of use.

\begin{table}[]
    \centering
    \caption{Zenodo repositories containing datasets and Python scripts related to this paper, including further documentation. Complete dataset exceeds 200GB. See Data Availability for terms and licences.}
    \label{tab:repositories}
    \scalebox{0.8}{\begin{tabular}{l|l}
    \hline
    \multicolumn{1}{c|}{\textbf{DOI}} & \multicolumn{1}{c}{\textbf{Repository content}} \\ \hline
    \rowcolor[rgb]{0.839,0.839,0.839}
    \url{10.5281/zenodo.13254469} & Raw ECMWF/ HRES (from 2019 to 2021) \\
    \url{10.5281/zenodo.13255978} & Raw ECMWF/ HRES (from 2022 to 2023) \\
    \rowcolor[rgb]{0.839,0.839,0.839} 
    \url{10.5281/zenodo.13255991} & Raw ECMWF/ ENS (from 2019 to 2020) \\
    \url{10.5281/zenodo.13256007} & Raw ECMWF/ ENS (2021) \\
    \rowcolor[rgb]{0.839,0.839,0.839}
    \url{10.5281/zenodo.13256010} & Raw ECMWF/ ENS (from 2022 to 2023) \\
    \url{10.5281/zenodo.13256014} & Raw wind power production data from \acrshort{bmra} \\
    \rowcolor[rgb]{0.839,0.839,0.839}
    \url{10.5281/zenodo.13309890} & Pre-processed BMRA data and scripts \\
    \url{10.5281/zenodo.13256016} & Produced wind power forecasts and evaluation \\
    \rowcolor[rgb]{0.839,0.839,0.839}
    \url{10.5281/zenodo.13309948} & Scripts used to produce forecasts and evaluation \\ \hline
    \end{tabular}}
\end{table}

\FloatBarrier
\bibliographystyle{elsarticle-num-names}
\bibliography{probW2P}

\end{document}